\documentclass[12pt,epsf]{article}

\usepackage{graphicx}
\usepackage{amsfonts,amssymb,amsmath}
\usepackage{graphicx}
\usepackage{latexsym}
\usepackage{bm}
\usepackage{cite}

\usepackage{color}

\newcommand{\be}{\begin{equation}}

\newcommand{\ee}{\end{equation}}

\newcommand{\ba}{\begin{array}}

\newcommand{\ea}{\end{array}}

\begin{document}
\begin{titlepage}
\vspace{.5in}
\begin{flushright}
CQUeST-2011-0455
\end{flushright}
\vspace{0.5cm}

\begin{center}
{\Large\bf Oscillating instanton solutions in curved space}\\
\vspace{.4in}

  {$\rm{Bum-Hoon \,\, Lee}^{\dag\S}$}\footnote{\it email:bhl@sogang.ac.kr}\,\,
  {$\rm{Chul \,\, H. \,\, Lee}^{\P}$}\footnote{\it
  email:chulhoon@hanyang.ac.kr}\,\,
  {$\rm{Wonwoo \,\, Lee}^{\S}$}\footnote{\it email:warrior@sogang.ac.kr}\, \,
  {$\rm{Changheon \,\, Oh}^{\P}$}\footnote{\it
  email:och0423@hanyang.ac.kr} \\

  {\small \dag \it Department of Physics and BK21 Division, Sogang University, Seoul 121-742,
  Korea}\\
  {\small \S \it Center for Quantum Spacetime, Sogang University, Seoul 121-742,
  Korea}\\
  {\small \P \it Department of Physics, Hanyang University, Seoul 133-791,
  Korea}\\

\vspace{.5in}
\end{center}
\begin{center}
{\large\bf Abstract}
\end{center}
\begin{center}
\begin{minipage}{4.75in}

{\small \,\,\,\, We investigate oscillating instanton
solutions of a self-gravitating scalar field between degenerate vacua. We show that there exist $O(4)$-symmetric oscillating solutions in a de Sitter background. The geometry of this solution is finite and preserves the $Z_{2}$ symmetry. The nontrivial solution corresponding to tunneling is possible only if the effect of gravity is taken into account. We present numerical solutions of this instanton, including the phase diagram of solutions in terms of the parameters of the present work and the variation of energy densities. Our solutions can be interpreted as solutions describing an instanton-induced domain wall or braneworld-like object rather than a kink-induced domain wall or braneworld. The oscillating instanton solutions have a thick wall and the solutions can be interpreted as a mechanism providing nucleation of the thick wall for topological inflation. We remark that $Z_{2}$ invariant solutions also exist in a flat and anti-de Sitter background, though the physical significance is not clear. }


\end{minipage}
\end{center}
\end{titlepage}

\newpage
\section{ Introduction \label{sec1}}

The interpolating solution between degenerate vacua in curved space can be the solution representing formation of a domain wall or a braneworld-like object with $Z_2$ symmetry if the thin-wall approximation scheme is used in the theory. Moreover, the solution after the analytic continuation expands, from an observer's point of view on the wall, without eating up bulk (inside and outside) spacetime. Can the braneworld or domain wall have a nontrivial internal structure? The structure can be made with oscillating instanton solutions. What is the meaning of this internal structure? Can the number of oscillations indicate certain criteria based on specific parameters? In order to get the answers to the above questions, we need to obtain oscillating solutions, interpret the meaning of the solutions, and construct a phase space of all allowed oscillating solutions.

In the absence of gravity, instantons are usually defined as solutions with a finite action to the classical field equations in Euclidean space obeying appropriate boundary conditions. The Yang-Mills instantons are also finite action solutions and have been studied extensively in gauge theories \cite{bpst, bl10} as well as in string theory \cite{ggp, vni}, and references therein. In gravitational theory, there are several kinds of instantons \cite{hghh}. Instantons should also be solutions to the Euclidean Einstein equations. Some instantons usually cannot be analytically continued to Lorentzian. Others can be analytically continued to Lorentzian. Instantons may be related to the semiclassical description of quantum gravity. However, this issue is beyond the scope of the present work and is therefore not being discussed here.

In this work, we consider tunneling phenomena in a double-well potential. Classically, a particle trapped in one well cannot penetrate through the potential barrier of that well, thus unaffected by the presence of the other well. However, if the central potential barrier is not infinitely high, there is tunneling between the two minima. The tunneling process is quantum mechanically described by the Euclidean solution obeying appropriate boundary conditions. The Euclidean solution interpolates between two different classical vacua. There exist two kinds of Euclidean solutions describing  tunneling phenomena in the double-well potential. One, which is for tunneling in an asymmetric double-well potential, corresponds to a bounce solution. The bounce solution describes the decay of a background vacuum state. The other, which is for tunneling in a symmetric double-well potential, corresponds to an instanton solution. The instanton solution, in this case, describes a general shift in the ground state energy of the classical vacuum due to the presence of an additional potential well, lifting the classical degeneracy \cite{colm02}.

The instanton solution in one dimension is equivalent to a static soliton in $(1+1)$ dimensions. Thus, the method for kink solution in two dimensions can be employed for studying the instanton solution in one dimension. The Euclidean equation of motion can be treated like a one-particle equation of motion with an evolution parameter playing the role of time in the inverted potential. A particle starts at the top of one hill at minus infinity in Euclidean time and arrives at the top of the other hill at plus infinity. The Euclidean action for the tunneling solution can be calculated easily using the fact that the Euclidean energy of an instanton is equal to zero in the system. This action is identical to the energy of a static solution of the $(1+1)$-dimensional soliton theory. In addition, the action has the same value as the action obtained in connection with the WKB approximation of the splitting in energies of the two lowest levels for the symmetric double-well potential. In semiclassical approximation, the action is dominated by classical configuration in evaluating the path integral.

For field theoretical solutions in $(3+1)$ dimensions, the Euclidean equation of motion for $O(4)$ symmetry has an additional term, which can be interpreted as a damping term. Thus, it is difficult to obtain the $O(4)$-symmetric instanton solution in this system when gravity is switched off. The damping term can be changed into the antidamping term not only in de Sitter (dS) space but also in both flat and anti-de Sitter (AdS) space when gravity is taken into account \cite{lllo}. The solutions have exact $Z_2$ symmetry and give rise to geometry of a finite size. Using a homogeneous and static scalar field, which has a constant value of $\Phi$ everywhere in Euclidean space, was shown to be impossible for the solution to have an infinite size due to the infinite cost of energy.

The bounce solution is related to the nucleation of a true (false) vacuum bubble describing decay of a background vacuum state. The process has been studied within various contexts for several decades. It was first investigated in Ref.\ \cite{voloshin} and developed in both flat \cite{col002} and curved spacetime \cite{bnu02, par02}. A homogeneous Euclidean configuration in which the scalar field jumps simultaneously onto the top of the potential barrier was investigated in Ref.\ \cite{hawking} and further studied in Ref.\ \cite{jst01}. As a special case of the true vacuum bubble, a vacuum bubble with a finite-sized background after nucleation was studied in Ref.\ \cite{bj05}. The decay of false monopoles with a gauge group was also studied using the thin-wall approximation \cite{kpy00}. The bubble or brane resulting from flux tunneling was  studied in a six-dimensional Einstein-Maxwell theory \cite{bsv00}.

The mechanism for nucleation of a false vacuum bubble in a true vacuum background has also been studied within various contexts. Nucleation of a large false vacuum bubble in dS space was obtained in Ref.\ \cite{kw04} and nucleation with a global monopole in Ref.\  \cite{yms}. The mechanism for nucleation of a small false vacuum bubble was obtained in the Einstein gravity with a nonminimally coupled scalar field \cite{wl01}, with Gauss-Bonnet term in Ref.\ \cite{koh01}, and using Brans-Dicke type theory \cite{kllly01}. The classification of vacuum bubbles including false vacuum bubbles in the dS background in the Einstein gravity was obtained in Ref.\ \cite{bl03}, in which the transition rate and the size of the instanton solution were evaluated in the space, as the limiting case of large true vacuum bubble or large false vacuum bubble.

The oscillating solution with $O(4)$ symmetry in dS space was first studied in Ref.\ \cite{hw000}, where the authors found the solution to oscillating scalar field $\Phi$ in fixed background. They adopted the fact that the role of damping term in the particle analogy picture can be changed into that of antidamping term in dS space if the evolution parameter exceeds half of a given range. The oscillation means that the field in their solutions oscillates back and forth between the two sides of the potential barrier. To obtain the general solutions including the effect of the backreaction, we will solve the coupled equations for the gravity and the scalar field simultaneously.

In the absence of gravity, the spectrum of small perturbations about the bounce solution has exactly one negative mode \cite{ccol}. When gravity is taken into account, the problem becomes more complex \cite{nemode}. On the other hand, the spectrum about the instanton solution between degenerate vacua has no negative mode. Thus we expect that the spectrum about our solutions will not have any negative modes.

The paper is organized as follows: in the next section we set up the basic framework for this work. We consider the tunneling process for the symmetric double-well potential and study the boundary conditions for our present work in detail. The boundary conditions analyzed in this work can be applicable to other cases including bounce solutions. In Sec.\ III, we present numerical oscillating instanton solutions by solving the coupled equations for the metric and the scalar field simultaneously. We show that there exist $O(4)$-symmetric oscillating solutions in dS background. We also examine the variation in the thickness of the wall and the energy density as the number of oscillations increased. Additionally, we examine numerically which solutions among oscillating solutions are relatively probable. The numerical solution is also possible in flat and AdS space as long as the local maximum value of the potential is positive. In Sec.\ IV, we analyze the properties of oscillating instanton solutions and construct the phase space of all our solutions in terms of the two parameters for this work. The phase space of solutions exhibits the behaviors that occur when the number of oscillations increases and indicates the regions where there are no solutions depending on the parameters. In Sec.\ V, we summarize and discuss our results. We interpret the physical meaning of our oscillating solutions with thick walls and discuss the probability of our solutions.

\section{The setup and the boundary conditions \label{sec2}}

The vacuum-to-vacuum transition amplitude called the generating functional for the Green's function can serve as a starting point for the nonperturbative treatment of the theory. In the path integral formalism, it can be visualized as the summation over all possible paths moving from the initial to the final state. The transition amplitude in the semiclassical approximation is dominated by those paths for which the Euclidean action difference is stationary. Thus, the tunneling amplitude can be evaluated in terms of the classical configuration and represented as $Ae^{-\Delta S}$ in this approximation, where the leading semiclassical exponent $\Delta S$ is the difference between the Euclidean action corresponding to a classical solution and the background action itself. The prefactor $A$ comes from the first order quantum correction \cite{colm02, ccol}.

Let us consider the following action:
\begin{equation}
S= \int_{\mathcal M} \sqrt{-g} d^4 x \left[ \frac{R}{2\kappa}
-\frac{1}{2}{\nabla^\alpha}\Phi {\nabla_\alpha}\Phi -U(\Phi)\right]
+ \oint_{\partial \mathcal M} \sqrt{-h} d^3 x \frac{K-K_o}{\kappa},
\label{f-action}
\end{equation}
where $g\equiv det g_{\mu\nu}$, $\kappa \equiv 8\pi G$, $R$ denotes the scalar curvature of the spacetime $\mathcal M$, $K$ and $K_o$ are the traces of the extrinsic curvatures of $\partial \mathcal M$ in the metric $g_{\mu\nu}$ and $\eta_{\mu\nu}$, respectively, and the
second term on the right-hand side is the boundary term \cite{York}. The gravitational field equations can be obtained properly from a variational principle with this boundary term. This term is also necessary to obtain the correct action.

We reconsider the tunneling process for the symmetric double-well potential in curved space similar to our previous work \cite{lllo}. The potential $U(\Phi)$, which represents the energy density of a homogeneous and static scalar field, has two degenerate minima
\begin{equation}
U(\Phi) = \frac{\lambda}{8} \left(\Phi^2 - \frac{\mu^2}{\lambda}
\right)^2 + U_{0}.   \label{pot01}
\end{equation}
The cosmological constant is given by $\Lambda=\kappa U_o$, hence the space will be dS, flat, or AdS depending on whether $U_{0} > 0$,
$U_{0} = 0$, or $U_{0} < 0$. We will make an attempt to obtain oscillating instanton solutions describing tunneling between degenerate vacua in these backgrounds.

To evaluate $\Delta S$ and show the existence of the solution, one has to take the analytic continuation to Euclidean space. We assume the $O(4)$ symmetry for both the geometry and the
scalar field as in Ref.\ \cite{bnu02}
\begin{equation}
ds^{2} = d\eta^{2} + \rho^{2}(\eta) \left[ d\chi^{2} + \sin^{2}\chi
\left( d\theta^{2} + \sin^{2}\theta d\phi^{2} \right) \right] .
\end{equation}
In this case $\Phi$ and $\rho$ depend only on $\eta$, and the Euclidean field equations for them can be written in the form:
\begin{equation}
\Phi'' + \frac{3\rho'}{\rho}\Phi'=\frac{dU}{d\Phi} \,\,\, {\rm and} \,\,\,
\rho'' = - \frac{\kappa}{3}\rho (\Phi'^2 +U), \label{erho}
\end{equation}
respectively and the Hamiltonian constraint is given by
\begin{equation}
\rho'^2 - 1 -
\frac{\kappa\rho^2}{3}\left(\frac{1}{2}\Phi'^{2}-U\right) = 0 .
\label{eqcon}
\end{equation}
In order to yield the solution the constraint requires a delicate balance among the terms. Otherwise the solution can provide qualitatively incorrect behavior \cite{berg}.

Now we have to consider the boundary conditions to solve Eqs.\
(\ref{erho}) and (\ref{eqcon}). There are two different methods. The first one is an initial value problem, in which we impose the initial conditions for the equations. For this to work, initial conditions are provided for the values of the fields $\rho$ and $\Phi$ or their derivatives $\rho'$ and $\Phi'$ at $\eta=0$ as follows:
\begin{equation}
\rho|_{\eta=0}=0, \,\,\,\,
\frac{d\rho}{d\eta}\Big|_{\eta=0}=1, \,\,\,\,  \Phi|_{\eta=0}=\Phi_o , \,\,\,\, {\rm and}\,\,\,\, \frac{d\Phi}{d\eta}\Big|_{\eta=0}=0, \label{ourbc-2}
\end{equation}
where the first condition means that the space including a solution is a geodesically complete space. The second condition stems from Eq.\ (\ref{eqcon}). The fourth condition $\Phi' = 0$ is due to the regularity condition as can be seen from the first equation in Eq.\ (\ref{erho}). However, the third condition for the initial value of $\Phi$ is not determined. One should find the initial value of $\Phi$ using the undershoot-overshoot procedure. This procedure particularly useful for the nonoscillating bounce solution, i.e. a one-crossing solution as we will explain about it in the next section. Some initial $\Phi_o$ will give the overshooting, in which the value of $\Phi$ at late $\eta$ value will go beyond $\mu/\sqrt{\lambda}$. Some other initial value of $\Phi_o$ will give the undershoot, in which the value of $\Phi$ at late $\eta$ does not reach $\mu/\sqrt{\lambda}$. Thus, the value $\Phi_o$ must be within an intermediate position between the undershoot and overshoot \cite{col002}. The existence of oscillating solutions is hidden within the undershoot as we will see in the next sections.
The other method is a boundary value problem or a two boundary value problem. We impose conditions specified at $\eta = 0$ and $\eta=\eta_{max}$. For this to work, we choose the values of the field $\rho$ and derivatives of the field $\Phi$ as follows:
\begin{equation}
\rho|_{\eta=0}=0 ,
\,\,\,\, \rho|_{\eta=\eta_{max}}=0, \,\,\,\,
\frac{d\Phi}{d\eta}\Big|_{\eta=0}=0, \,\,\,\, {\rm and}\,\,\,\,
\frac{d\Phi}{d\eta}\Big|_{\eta=\eta_{max}}=0. \label{ourbc-3}
\end{equation}
$\eta_{max}$ is the maximum value of $\eta$ and will have a finite value. These conditions are useful for obtaining solutions with $Z_2$ symmetry.

In order to solve the Euclidean field Eqs.\ (\ref{erho}) and
(\ref{eqcon}) numerically, we first rewrite the equations in terms of dimensionless variables as in Ref \cite{wl01}:
\begin{equation}
\frac{\lambda U(\Phi)}{\mu^{4}} = \tilde{U}(\tilde{\Phi}), \quad
\frac{\lambda \Phi^{2}}{\mu^{2}} = \tilde{\Phi}^{2}, \quad \mu\eta =
\tilde{\eta}, \quad \mu\rho = \tilde{\rho}, \quad {\rm and} \quad
\frac{\mu^{2}}{\lambda}\kappa = \tilde{\kappa}.
\end{equation}
These variables give
\begin{equation}
\tilde{U}(\tilde{\Phi}) = \frac{1}{8} \left( \tilde{\Phi}^{2} - 1
\right)^{2} + \tilde{U}_{0},
\end{equation}
and the Euclidean field equations for $\Phi$ and $\rho$ are rewritten as:
\begin{equation}
\tilde{\Phi}'' + \frac{3\tilde{\rho}'}{\tilde{\rho}}\tilde{\Phi}'
=
\frac{d\tilde{U}}{d\tilde{\Phi}} \,\,\, {\rm and} \,\,\,
\tilde{\rho}'' = - \frac{\tilde{\kappa}}{3}\tilde{\rho} (\tilde{\Phi}'^2
+\tilde{U}), \label{eqrho-change}
\end{equation}
respectively. The Hamiltonian constraint is given by
\begin{equation}
\tilde{\rho}'^{2} - 1 - \frac{\tilde{\kappa} \tilde{\rho}^{2}}{3}
\left( \frac{1}{2} \tilde{\Phi}'^{2} - \tilde{U} \right) = 0.
\label{hamconst}
\end{equation}

We now make comments on the boundary conditions more precisely to solve Eqs.\ (\ref{eqrho-change}) and (\ref{hamconst}) numerically. As we already mentioned in the boundary value problem, there are two kinds of conditions. The relaxation method is one of the methods employed to solve a two boundary value problem. However we do not know the exact value of $\tilde{\eta}_{max}$ in our case, thus we can not impose the boundary condition at $\tilde{\eta}=\tilde{\eta}_{max}$. In this work, we employ the shooting method using the adaptive step size Runge-Kutta as an initial value problem as done in Ref. \cite{NR}. For this procedure we choose the initial values of $\tilde{\Phi}(\tilde{\eta}_{initial})$, $\tilde{\Phi}'(\tilde{\eta}_{initial})$, $\tilde{\rho}(\tilde{\eta}_{initial})$, and $\tilde{\rho}'(\tilde{\eta}_{initial})$ at $\tilde{\eta}=\tilde{\eta}_{initial}$ as follows:
\begin{eqnarray}
\tilde{\Phi}(\tilde{\eta}_{initial}) &\sim& \tilde{\Phi}_{0} +
\frac{\epsilon^2}{16}\tilde{\Phi}_{0}\left( \tilde{\Phi}_{0}^2-1
\right) +
\frac{\epsilon^{3}}{48}\left( 3\tilde{\Phi}_{0}^{2} - 1 \right),  \nonumber \\
\tilde{\Phi}'(\tilde{\eta}_{initial}) &\sim& \frac{\epsilon}{8}
\tilde{\Phi}_{0} \left( \tilde{\Phi}_{0}^{2} - 1 \right) +
\frac{\epsilon^{2}}{16}\left( 3\tilde{\Phi}_{0}^2
- 1 \right),  \label{nbcon2} \\
\tilde{\rho}(\tilde{\eta}_{initial}) &\sim& \epsilon,  \nonumber \\
\tilde{\rho}'(\tilde{\eta}_{initial}) &\sim& 1 , \nonumber
\end{eqnarray}
where $\tilde{\eta}_{initial} = 0+\epsilon$ and $\epsilon \ll 1$. If we find the initial value $\tilde{\Phi}_{0}$, the other conditions are given by Eqs.\ (\ref{nbcon2}).  Furthermore we impose additional conditions implicitly. To avoid a singular solution at $\tilde{\eta} = \tilde{\eta}_{max}$ in Eq.\ (\ref{eqrho-change}) and to demand the $Z_2$ symmetry, the conditions $d\tilde{\Phi}/{d\tilde{\eta}} \rightarrow 0$ and $\tilde{\rho}\rightarrow 0$ as $\tilde{\eta}\rightarrow \tilde{\eta}_{max}$ are needed. In this work, we require that the value of $d\tilde{\Phi}/{d\tilde{\eta}}$ goes to a value smaller than $10^{-6}$ as $\tilde{\eta} \rightarrow \tilde{\eta}_{max}$, as the exact value of $\tilde{\eta}_{max}$ is not known.

We will examine the energy density of the oscillating region of the solution to observe a domain wall structure. The action for the one-dimensional instanton has the following form
\begin{equation}
s_o= \int^{\infty}_{-\infty} \left[\frac{1}{2}\Phi'^2 +U \right] d\tau = \frac{2\mu^3}{3\lambda}, \nonumber
\end{equation}
where we take the potential similar to the one in Eq.\ (\ref{pot01}). To obtain a solution with finite action, it is required that $\Phi' \rightarrow 0$ and $U \rightarrow 0$ as $\tau \rightarrow \pm\infty$. This action is equal to the surface tension or the surface energy density of the wall \cite{col002}. When gravity is taken into account, the situation is more complicated. To simplify things, we only consider the Euclidean action of the bulk part in Eq.\  (\ref{f-action}) to get,
\begin{equation}
S_E= \int_{\mathcal M} \sqrt{g_E} d^4 x_E \left[ -\frac{R_E}{2\kappa} +\frac{1}{2}\Phi'^2 +U \right] =  2\pi^2 \int \rho^3 d\eta [-U], \label{euclac}
\end{equation}
where $R_E =6[1/\rho^2 - \rho'^2/\rho^2 - \rho''/\rho]$ and we used Eqs.\ (\ref{erho}) and (\ref{eqcon}) to arrive at this. Thus the contributions coming from the geometry and kinetic energy in the Euclidean action are included to be the potential effectively. On the other hand, the action in Ref.\ \cite{bnu02} can be rewritten as
\begin{equation}
S_E= -\frac{12\pi^2}{\kappa} \int \rho d\eta \left( 1- \frac{\kappa \rho^2 U}{3}  \right),
\end{equation}
where the authors used integration by parts. Since they were only interested in the difference between the actions of the two solutions that agree at infinity, the surface term evolving from the integration by parts is regarded as unimportant. The minus sign appeared in the action is related to the fact that the Euclidean action for Einstein gravity is not bounded from below, which is known as the conformal factor problem in Euclidean quantum gravity \cite{ghp000}. However, this issue is not central to our problem and therefore we will not discuss this issue furthermore in the present paper. In this work, we will examine the relative probability among the solutions. We employ Eq.\ (\ref{euclac}) for the present work. The volume energy density has the following form
\begin{equation}
\xi \equiv -\mathcal{H} = -\left[ -\frac{R_E}{2\kappa} +
\frac{1}{2}{\Phi}'^{2} + U \right] = U. \label{voden}
\end{equation}
We will examine the change in the density with respect to the evolution parameter $\eta$ in the next section.

\section{Oscillating instanton solutions between degenerate
vacua} \label{sec3}

In Ref.\ \cite{hw000}, the authors found an oscillating solution between the dS-dS vacuum states. They have shown that there exist two kinds of oscillating solutions, one is endowed with an asymmetric double-well potential while the other is with a symmetric potential. They showed how does the maximum allowed number $n_{max}$ depend on the parameters of the theory, in which $n$ denotes the crossing number of the potential barrier by the oscillating solutions. However, they solved the equation for the scalar field in the fixed dS background. This is a good approximation only for $U_o \gg U(\Phi)-U_o $ \cite{hw000}. In this paper, we will solve the coupled equations for the metric and the scalar field simultaneously for any arbitrary vacuum energy $U_o$. In the first place, we present the details of oscillating instanton solutions including numerical solutions and the variation of the energy density to see the structure of the domain wall and the oscillating behavior depending on $\tilde{\kappa}$ and $\tilde{\kappa} \tilde{U}_o$ in dS space with $U_o >0$. Then, we present oscillating solutions in flat space with $U_o = 0$. Finally, we present oscillating solutions in the AdS space with $U_o < 0$.

\begin{figure}[t]
\begin{center}
\includegraphics[width=2.0in]{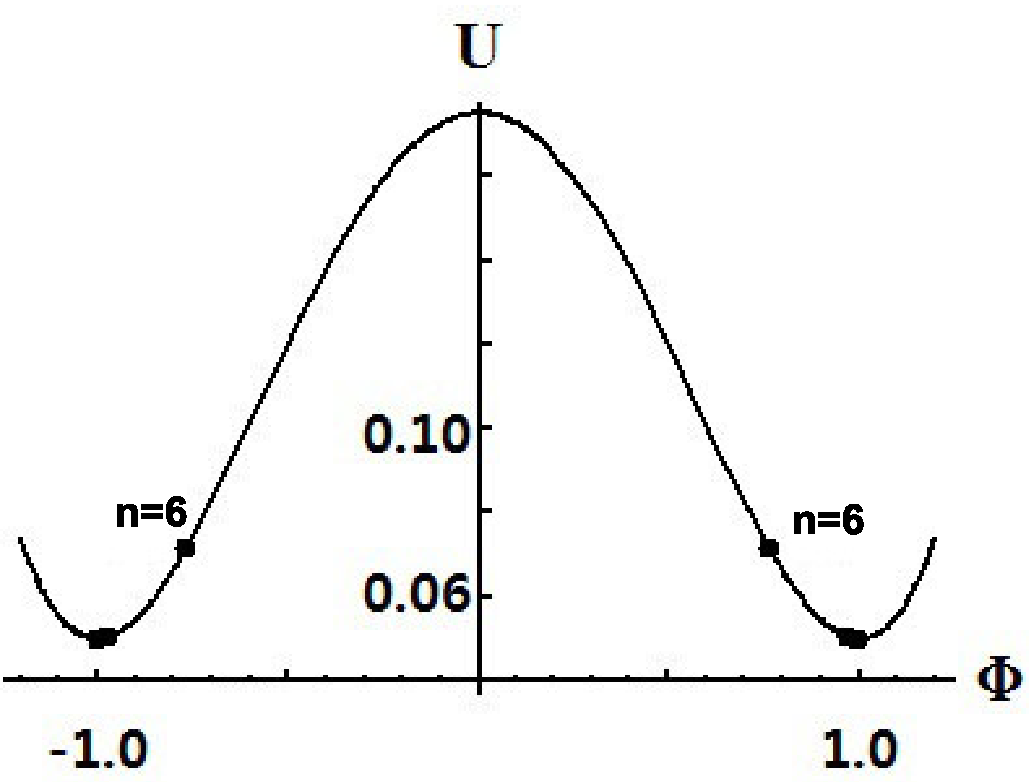}
\includegraphics[width=2.0in]{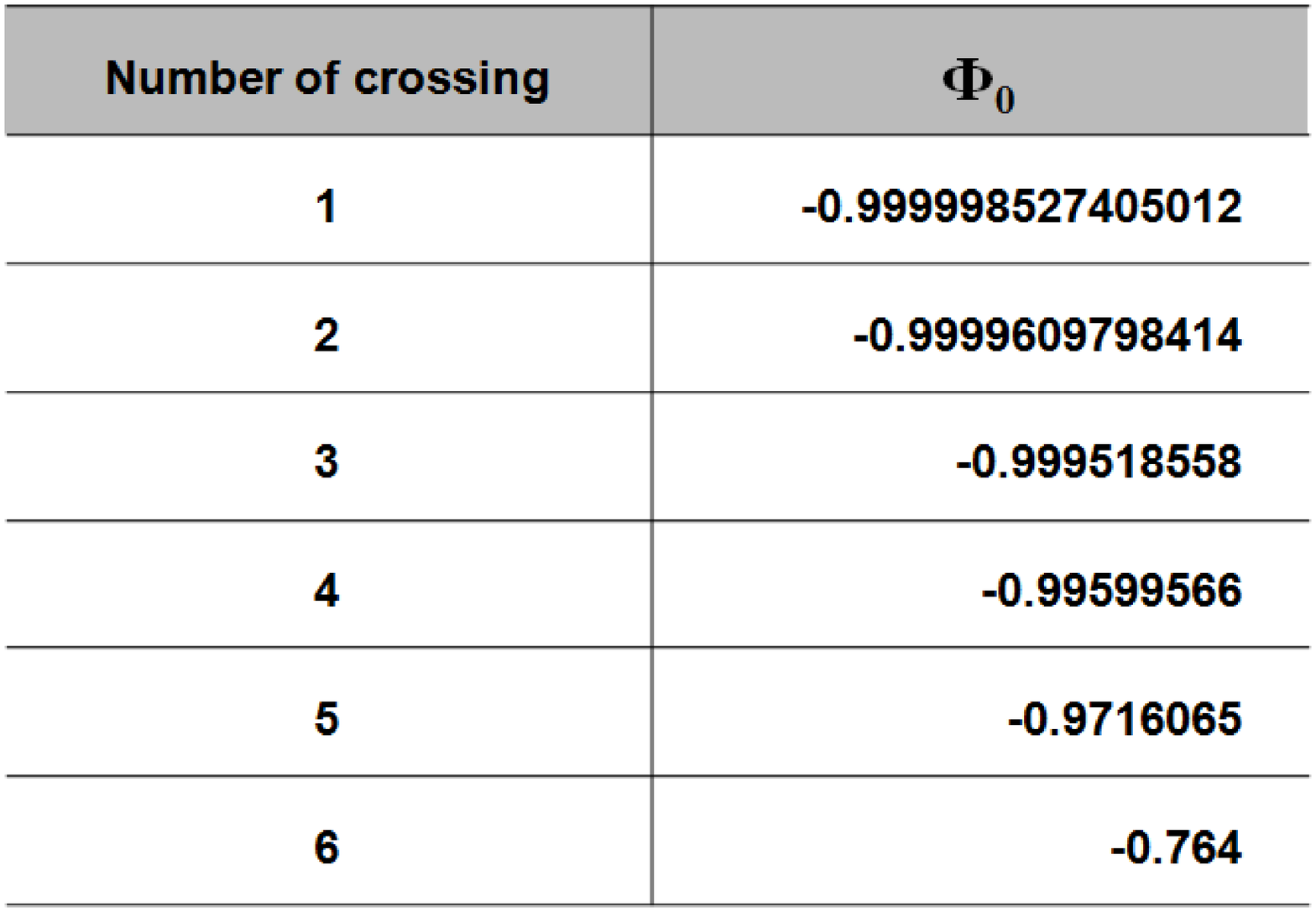}
\includegraphics[width=2.0in]{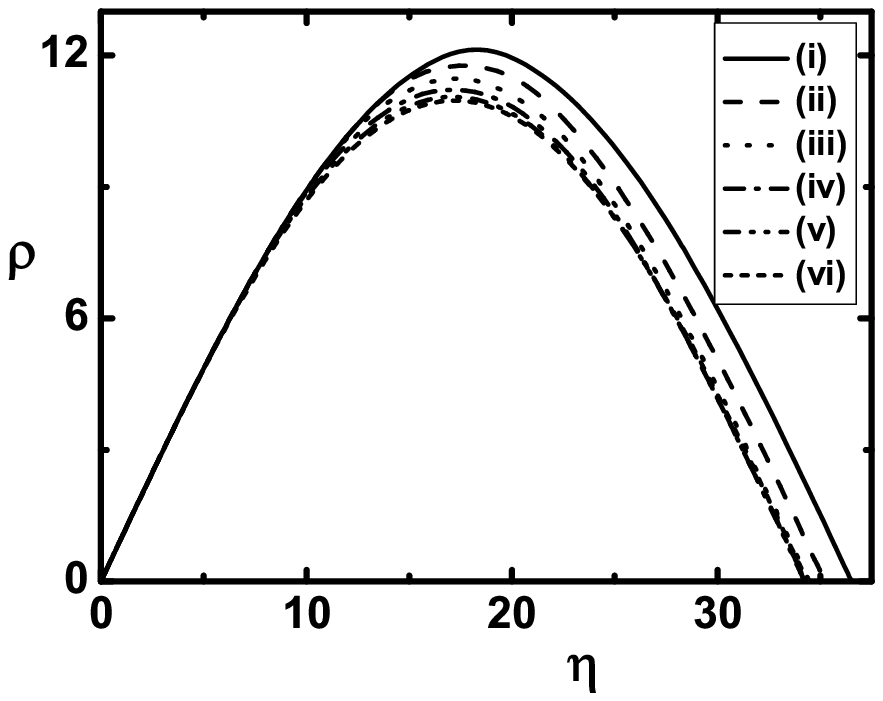}
\includegraphics[width=2.0in]{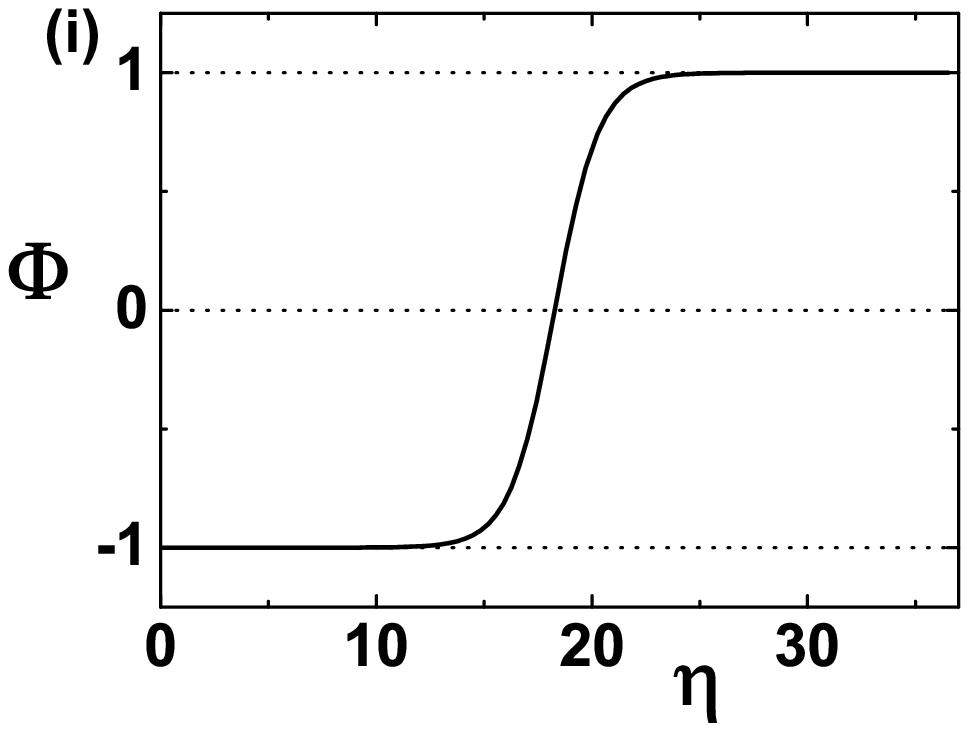}
\includegraphics[width=2.0in]{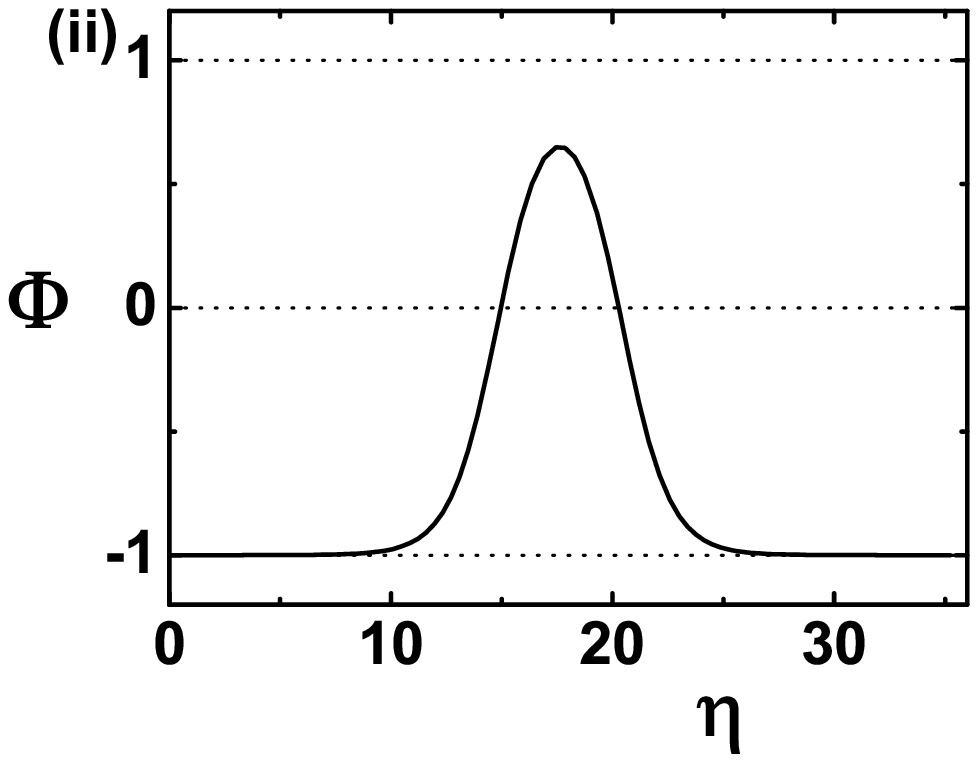}
\includegraphics[width=2.0in]{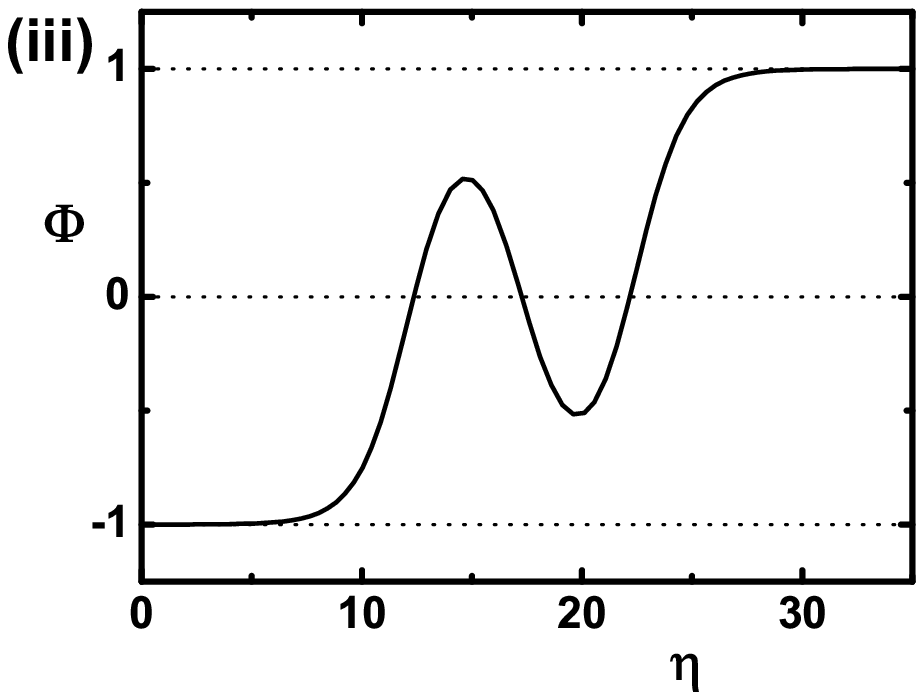}
\includegraphics[width=2.0in]{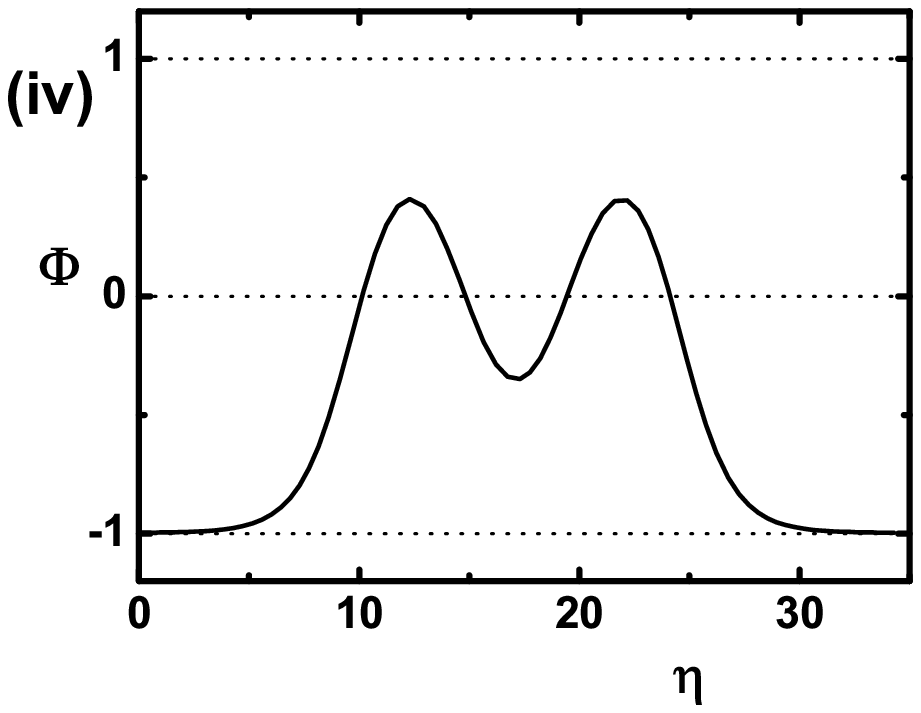}
\includegraphics[width=2.0in]{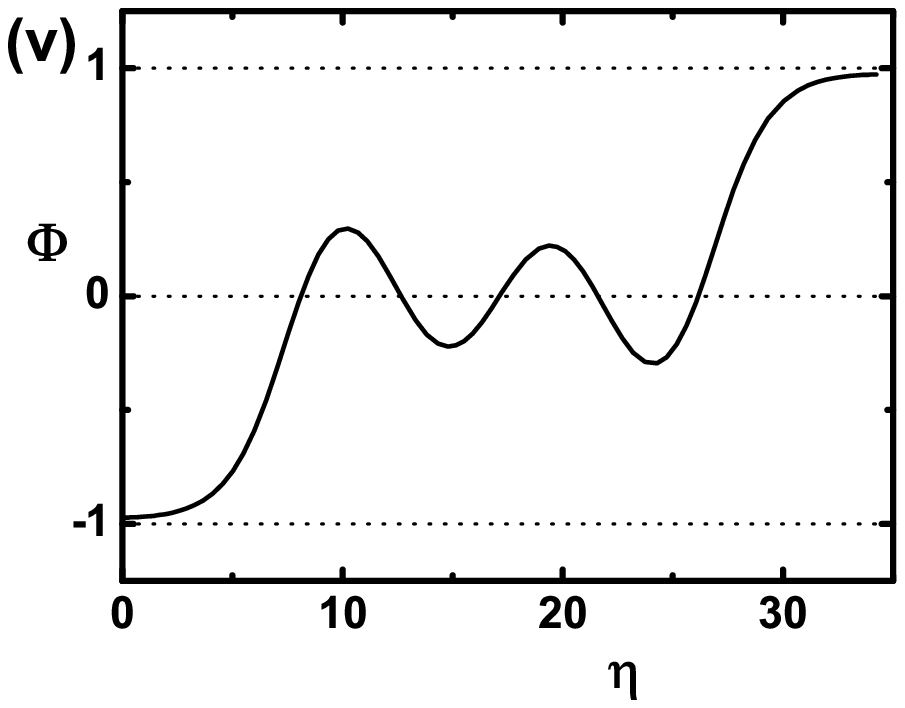}
\includegraphics[width=2.0in]{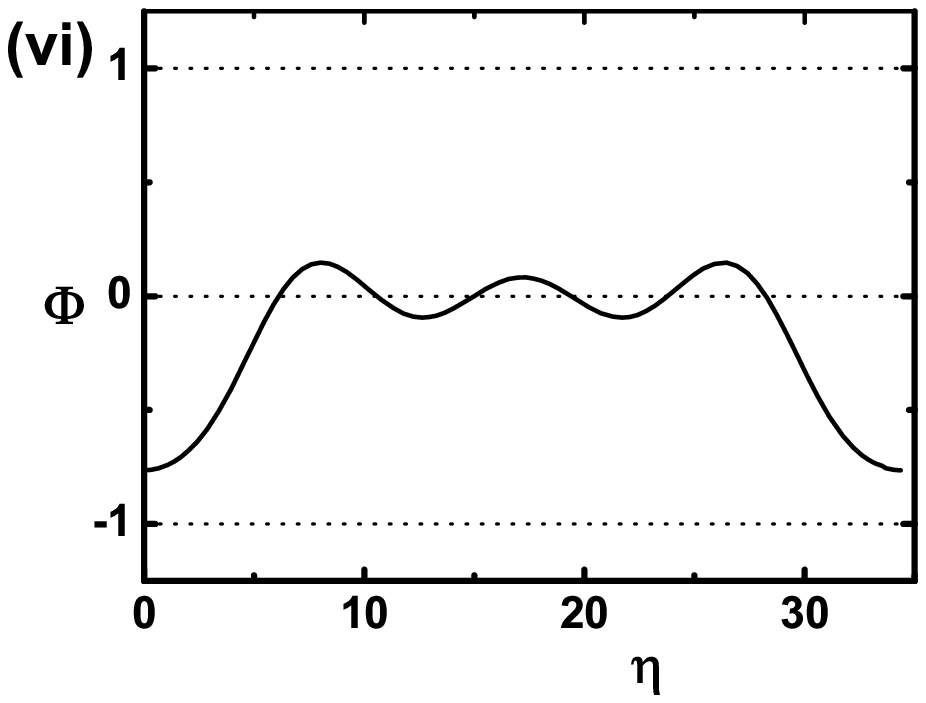}
\end{center}
\caption{\footnotesize{The numerical solutions represent oscillating instanton solutions between dS-dS degenerate vacua. }} \label{fig:fig01}
\end{figure}

\begin{figure}[t]
\begin{center}
\includegraphics[width=2.0in]{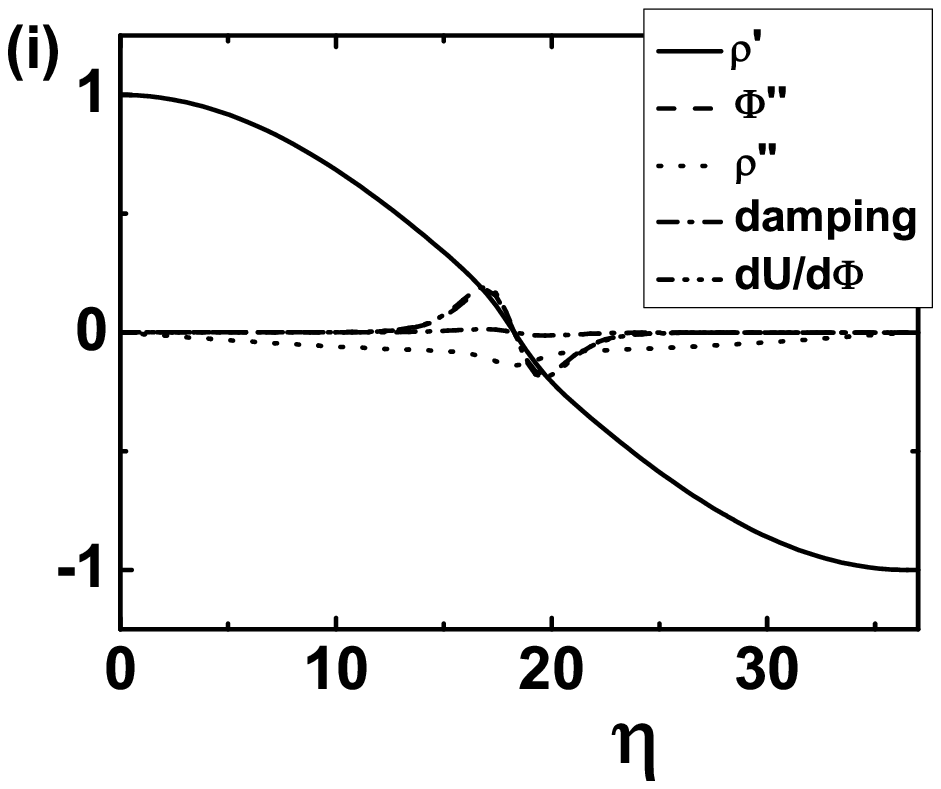}
\includegraphics[width=2.0in]{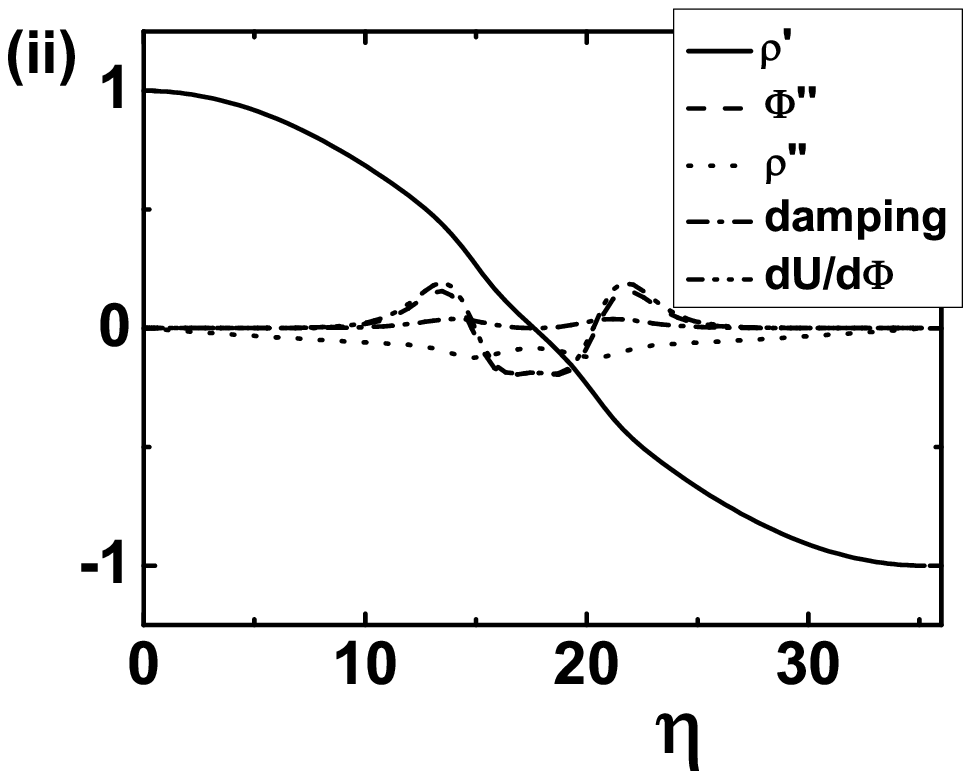}
\includegraphics[width=2.0in]{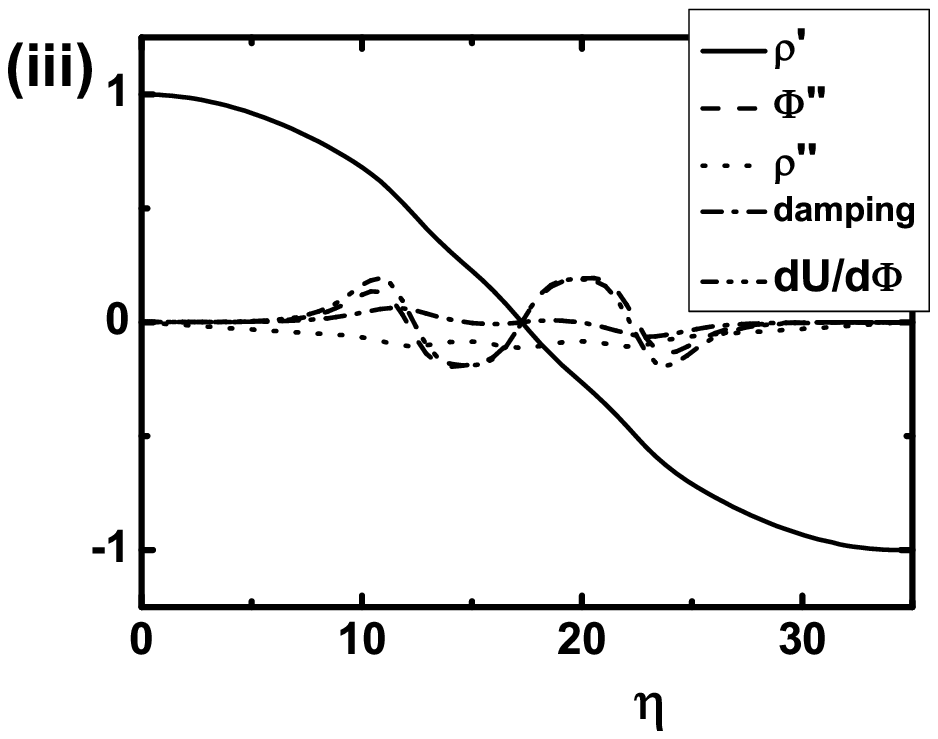}
\includegraphics[width=2.0in]{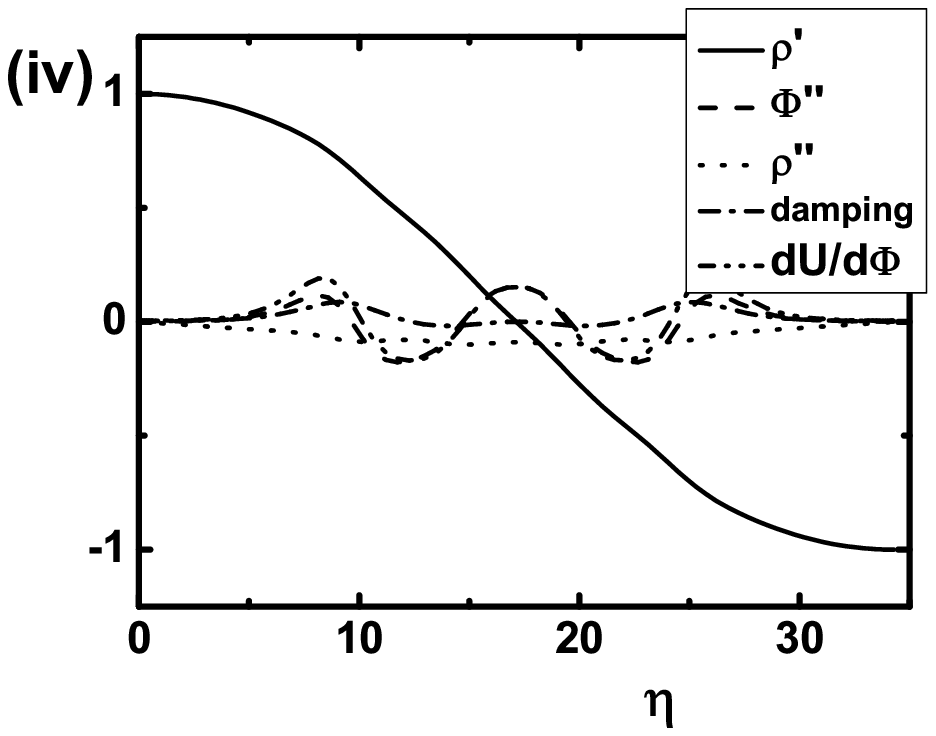}
\includegraphics[width=2.0in]{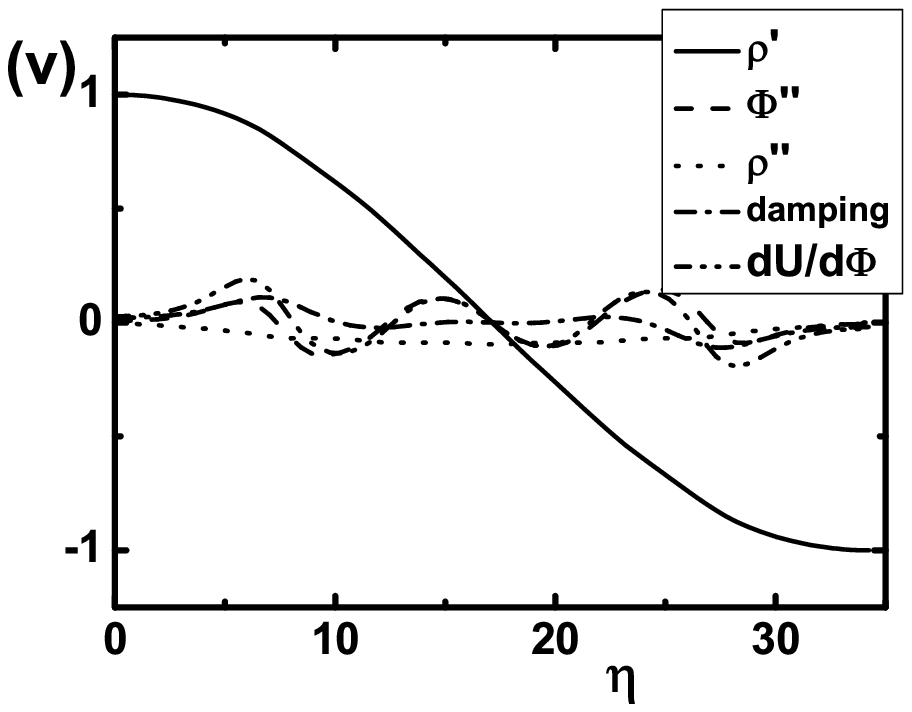}
\includegraphics[width=2.0in]{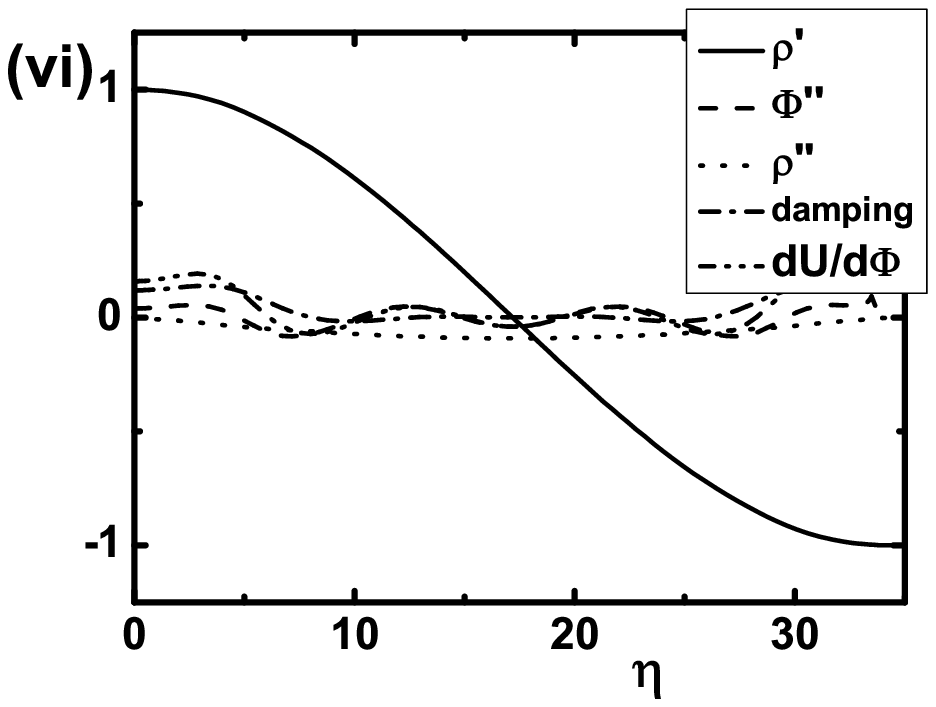}
\end{center}
\caption{\footnotesize{Variation of terms in equations of motion
between dS-dS degenerate vacua.}} \label{fig:fig02}
\end{figure}

Figure \ref{fig:fig01} shows oscillating instanton solutions representing tunneling starting from left vacuum state in dS-dS degenerate vacua. The first figure illustrates the potential, in which the number $n$ denotes the number of the crossing or the number of oscillations. We take $\tilde{U}_o = 0.5$ and $\tilde{\kappa} = 0.04$ for all the cases. The second figure illustrates the  initial point $\tilde{\Phi}_o$ for each number of oscillations. As expected, the number of oscillations increases as the initial point, $\tilde{\Phi}(\tilde{\eta}_{initial})=\tilde{\Phi}_o$, moves away from the vacuum state. The third figure illustrates the solutions of $\tilde{\rho}$. The solution of $\tilde{\rho}$ is $\sqrt{\frac{3}{\tilde{\kappa}\tilde{U}_o}} \sin\sqrt{\frac{\tilde{\kappa}\tilde{U}_o}{3}}\eta$ in fixed dS space. Thus, the graph of a sine type function near the vacuum states indicates dS space. We can see that the size of the geometry with such a solution decreases as the number of crossing increases because the period of the evolution parameter $\tilde{\eta}_{max}$ decreases as the starting point moves away from the vacuum state. Figure (i) illustrates the one-crossing solution, $n=1$, of the field $\tilde{\Phi}$. The one-crossing solution corresponds to the instanton solutions with the $O(4)$ symmetry between the degenerate vacua \cite{lllo}, for which it was shown that the solutions exist not only in the dS space but also in the flat and AdS space if the local maximum value of the potential is positive. Figure (ii) illustrates the two-crossing solution, $n=2$. The two-crossing solution, $n=2$, was considered as a type of the double-bounce solution or anti-double-bounce solution \cite{boulin}, in which the authors interpreted the double-bounce solution as the spontaneous pair-creation of true vacuum bubbles separated by a wall. However, the solution in this paper is quite different from the double-bounce solution found in Ref.\ \cite{boulin}. Since our solution of $\tilde{\Phi}$ does not asymptotically approach the other vacuum state, it is difficult to interpret our solution as the double-instanton solution or the spontaneous pair-creation of instanton solutions. Figures (iii)-(vi) illustrate the $n$-crossing solution, for $n=3, 4, 5, 6$. The maximum number $n_{max} = 6$ depends on the parameters $\tilde{U}_o$ and $\tilde{\kappa}$. In other words, there is no solution for $n> n_{max}$ with $\tilde{U}_o = 0.5$ and $\tilde{\kappa} = 0.04$. Figures (i), (iii), and (v) illustrate the tunneling starting from the left vacuum state to the right vacuum state. Figures (ii), (iv), and (vi) illustrate solutions going back to the starting point after oscillations. The maximum number of oscillations is determined by the parameters $\tilde{\kappa}$ and $\tilde{U}_o$ as was observed in Ref.\ \cite{hw000}.

\begin{figure}[t]
\begin{center}
\includegraphics[width=2.0in]{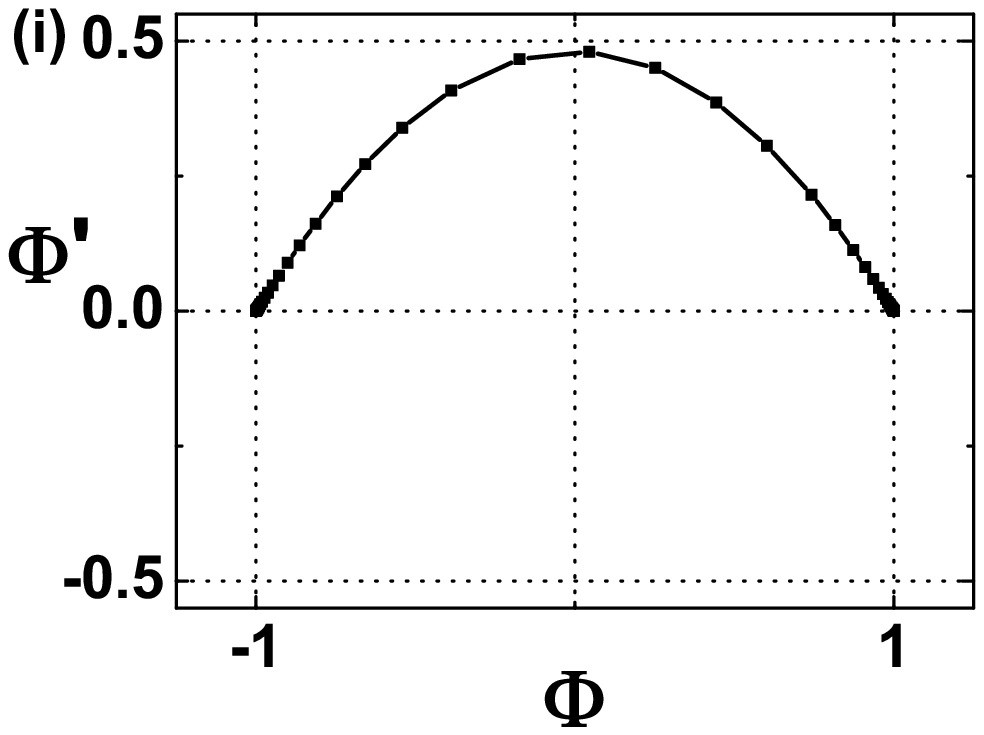}
\includegraphics[width=2.0in]{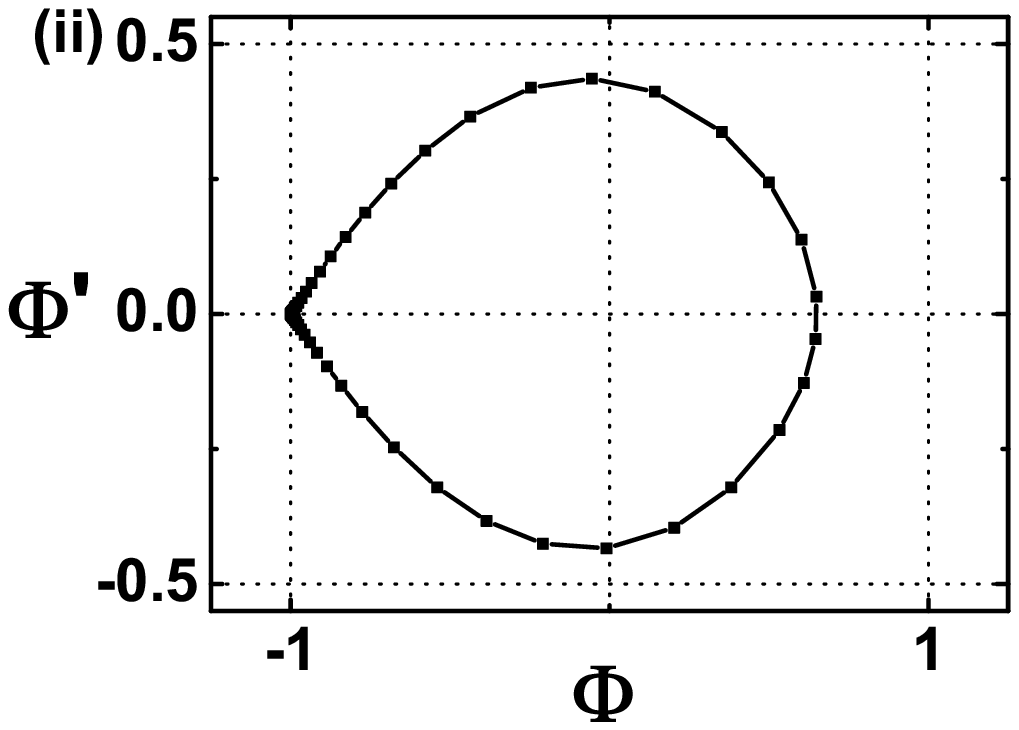}
\includegraphics[width=2.0in]{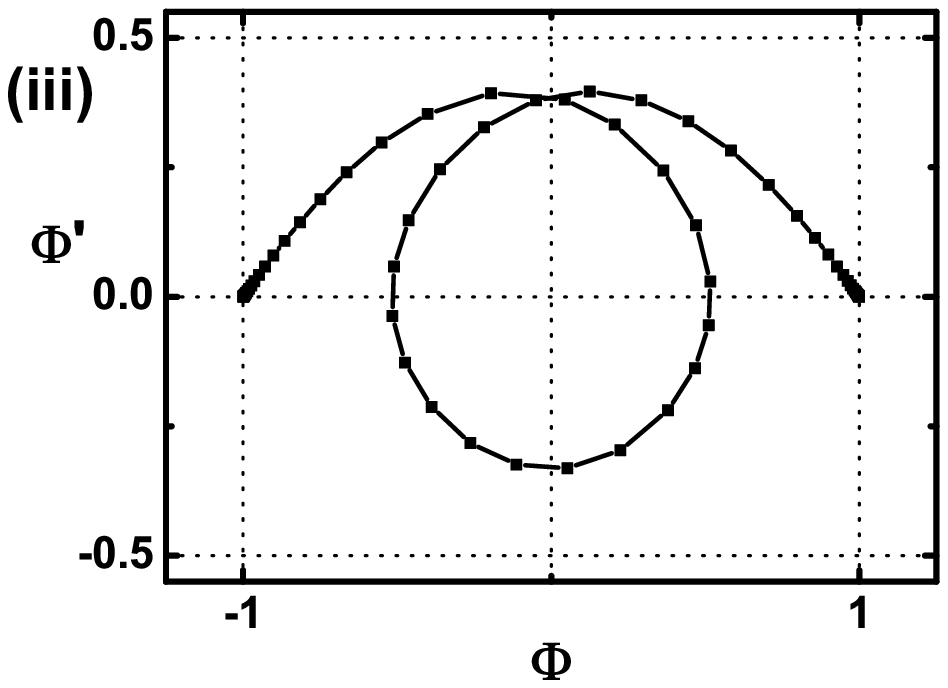}
\includegraphics[width=2.0in]{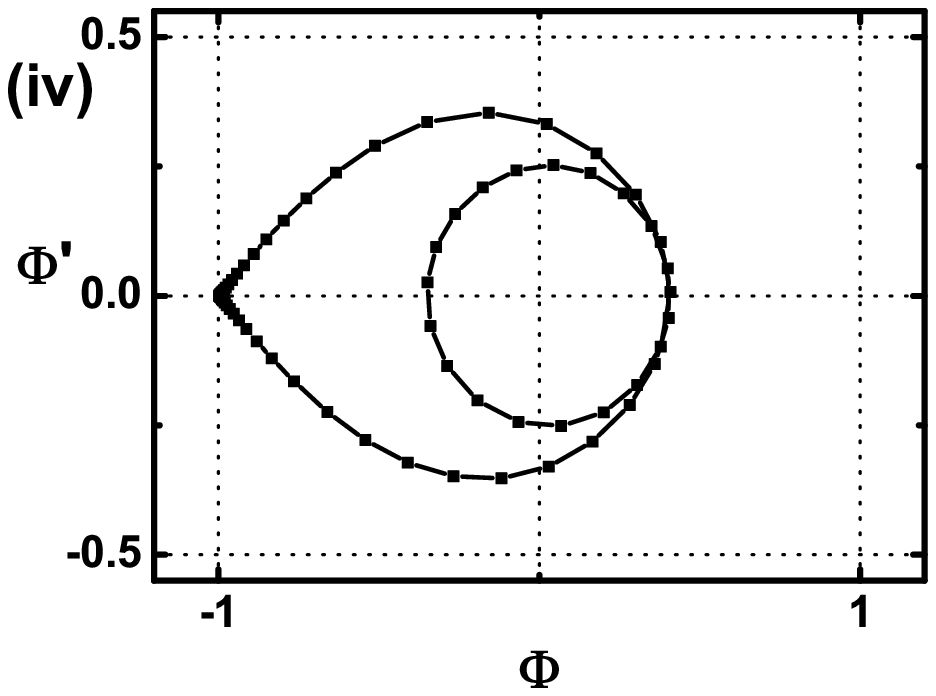}
\includegraphics[width=2.0in]{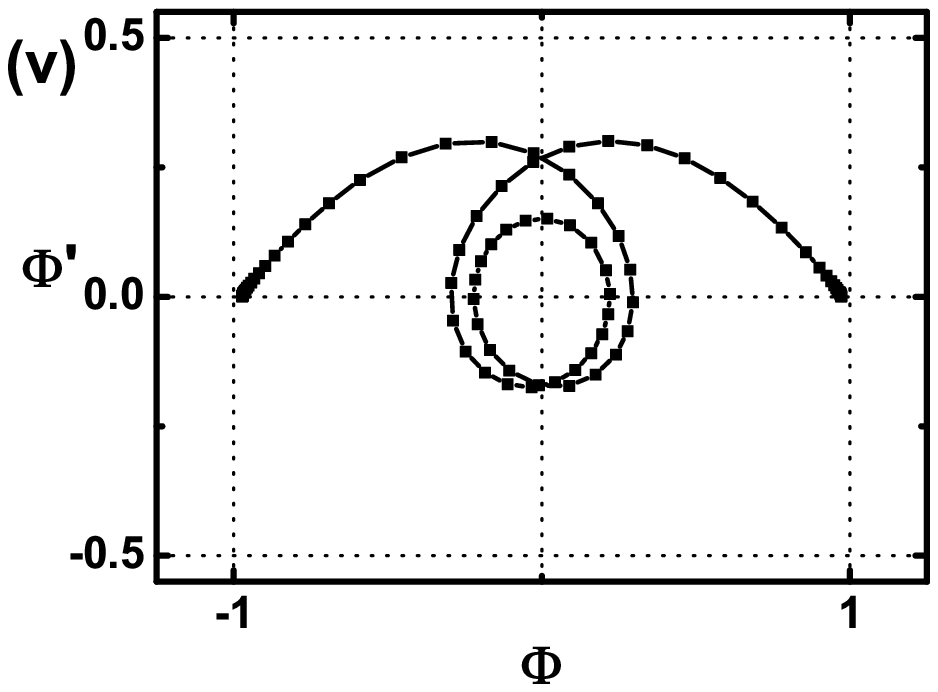}
\includegraphics[width=2.0in]{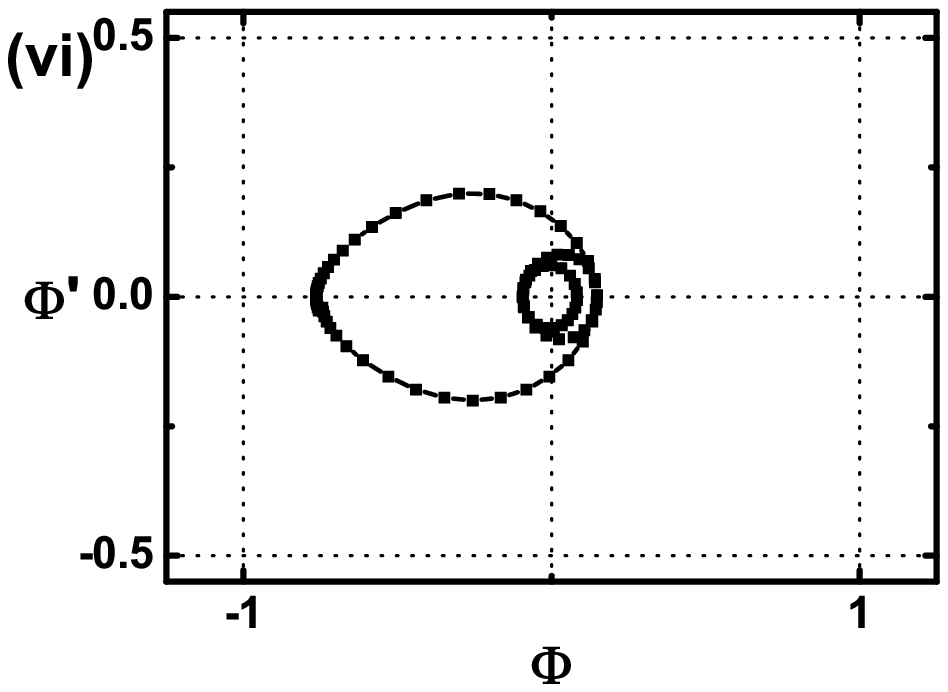}
\end{center}
\caption{\footnotesize{The behavior of the solutions in the $\Phi$-$\Phi'$ plane using the phase diagram method. This case belongs to the tunneling between dS-dS degenerate vacua.}} \label{fig:fig03}
\end{figure}

Figure \ref{fig:fig02} shows the variation of terms, $\tilde{\rho}'$, $\tilde{\Phi}'$, $\tilde{\rho}''$, $\frac{3\tilde{\rho}'} {\tilde{\rho}}\tilde{\Phi}'$, and $d\tilde{U}/d\tilde{\Phi}$, with respect to $\tilde{\eta}$ appearing in Eqs.\ (\ref{eqrho-change}) and (\ref{hamconst}). In figures (i) - (vi), we see that the change of sign of $\tilde{\rho}'$ from positive to negative occurs at the half period due to the $Z_2$ symmetry. The value of $\tilde{\rho}'$ at the initial and the final value of $\tilde{\eta}$ means there is a dS space at that point. The value of $\rho'$ spans from $-1$ to $1$ in all figures. The transition region of $\tilde{\rho}'$ means the rolling duration in the inverted potential. In this region, all other terms have dynamical behavior. The value of $\tilde{\Phi}''$ representing an acceleration of the particle in the inverted potential increases, decreases, and becomes zero at the half period. The graph is odd function. The value of $\tilde{\rho}''$ is always negative or zero as an even function according to Eq.\ (\ref{eqrho-change}). The damping term also increases, decreases, and becomes zero as an odd function. The term $d\tilde{U}/d\tilde{\Phi}$ has got the same property. The figures (i), (iii), and (v) representing tunneling show that $\tilde{\rho}'$, $\tilde{\Phi}''$, damping term, and $d\tilde{U}/d\tilde{\Phi}$ change their sign simultaneously at the half period. While the figures (ii), (iv), and (vi) representing solutions going back to the starting point show that only $\tilde{\rho}'$ change its sign at the half period. All of the behaviors represented in each figure can be well understood bearing $Z_2$ symmetry in mind. The initial and final regions of $\tilde{\rho}'$ in each figure exhibit cosine type function as the solution near the vacuum states indicates dS space.

The behavior of the solutions in the $\tilde{\Phi}(\tilde{\eta})$-$\tilde{\Phi}'(\tilde{\eta})$ plane using the phase diagram method is shown in Fig.\ \ref{fig:fig03}. The figure (i) illustrates the phase diagram of a one-crossing solution, in which the trajectory is restricted to the upper half region of the diagram. It is the turning point from the damping phase to the antidamping phase when $\tilde{\Phi}'$ takes the maximum value and $\tilde{\Phi}$ attains the first zero. The value of $\tilde{\Phi}$ spans from $-1$ to $+1$ and $\tilde{\Phi}'$ from zero via maximum value to $\thicksim 0.47$, to zero with symmetry about the $y$ axis. The figure (ii) illustrates the diagram of a two-crossing solution, in which the trajectory does not reach the opposite point $\tilde{\Phi}=1$ but return to the starting point of $\tilde{\Phi}=-1$. When the trajectory goes back, $\tilde{\Phi}'$ is negative with a symmetry about the $x$ axis. It is the turning point from the damping phase to the antidamping phase when $\tilde{\Phi}'$ reaches the second zero and $\tilde{\Phi}$ takes the positive value. The figure (iii) illustrates the diagram of a three-crossing solution. It is the turning point where $\tilde{\Phi}'$ takes the negative maximum value and $\tilde{\Phi}$ attains the second zero. The figure (iv) illustrates the diagram of a four-crossing solution. It is the turning point when $\tilde{\Phi}'$ reaches the third zero and $\tilde{\Phi}$ takes the negative value. The figure (v) illustrates the diagram of a five-crossing solution. It is the turning point when $\tilde{\Phi}'$ takes the positive value and $\tilde{\Phi}$ attains the second zero. The figure (vi) illustrates the diagram of a six-crossing solution. It is the turning point when $\tilde{\Phi}'$ reaches the third zero and $\tilde{\Phi}$ takes the positive value. The figures (i), (iii), and (v) have a symmetry about the $y$ axis, whereas figures (ii), (iv), and (vi) have a symmetry about the $x$ axis. The maximum value of $\tilde{\Phi}'$ decreases as the number of crossing increases.

\begin{figure}[t]
\begin{center}
\includegraphics[width=2.0in]{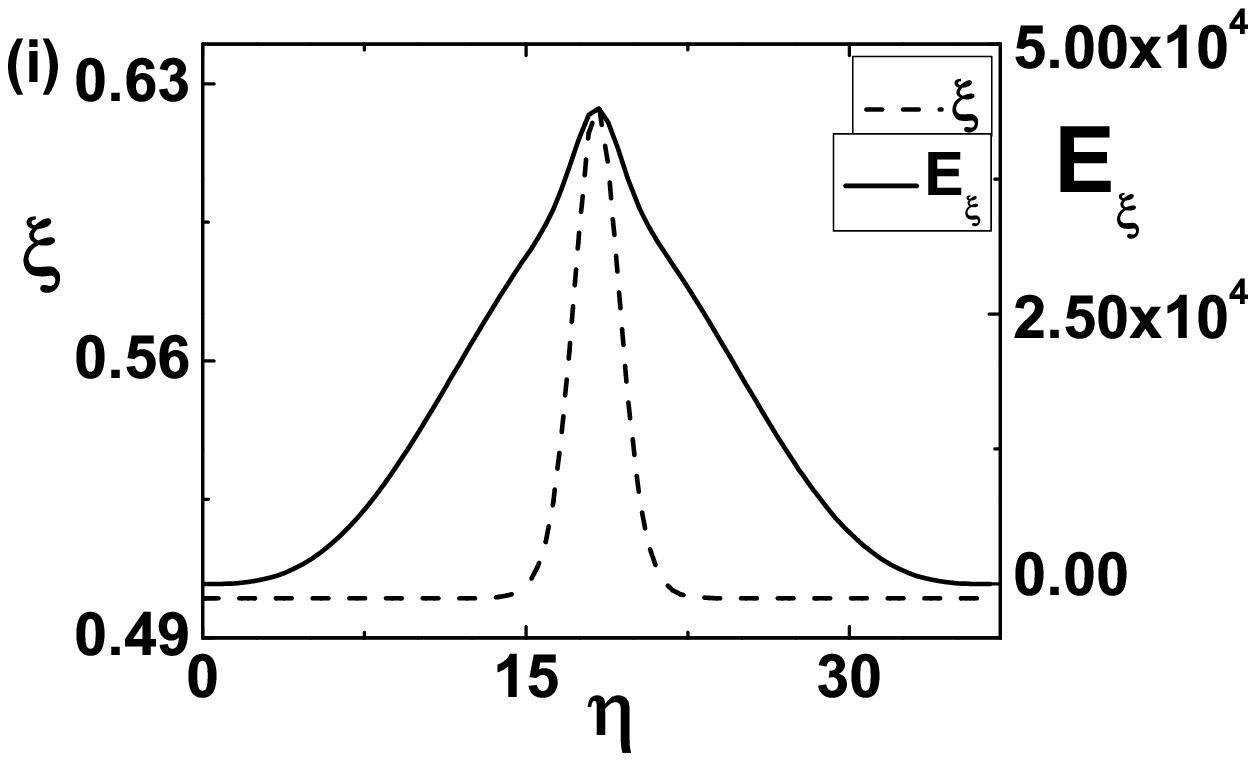}
\includegraphics[width=2.0in]{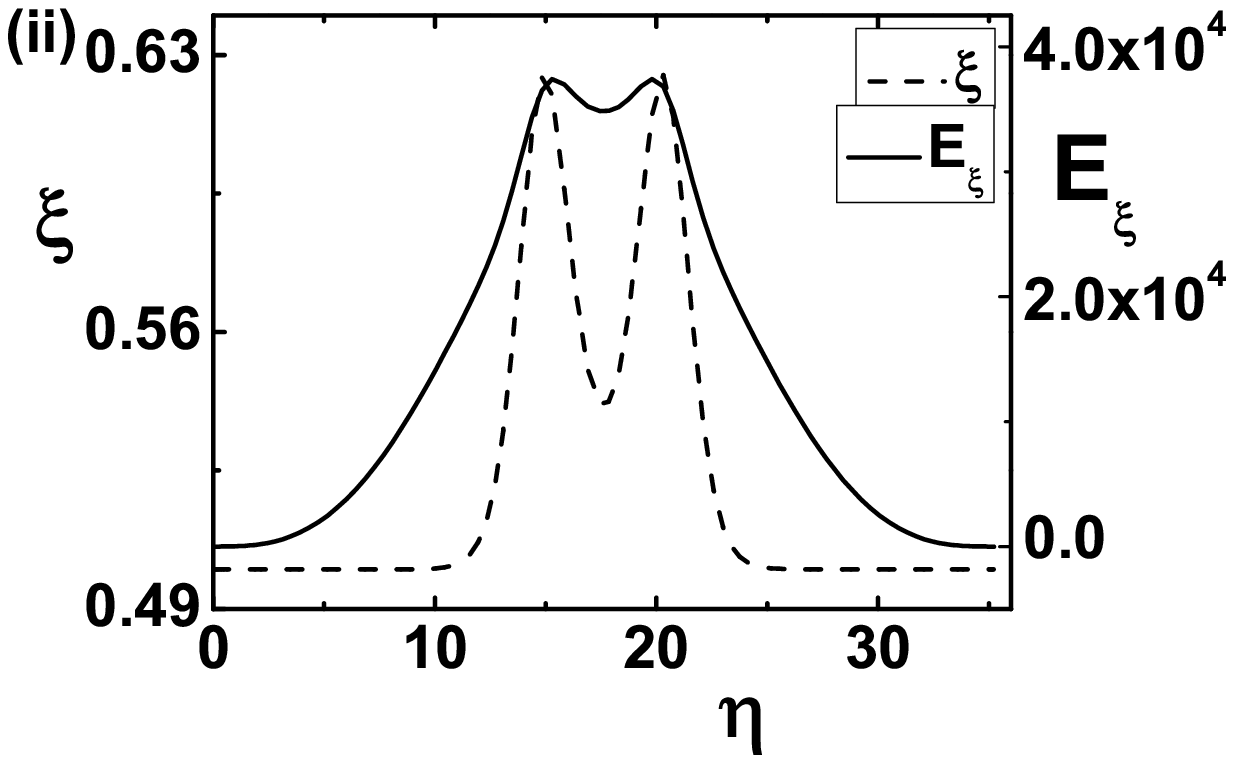}
\includegraphics[width=2.0in]{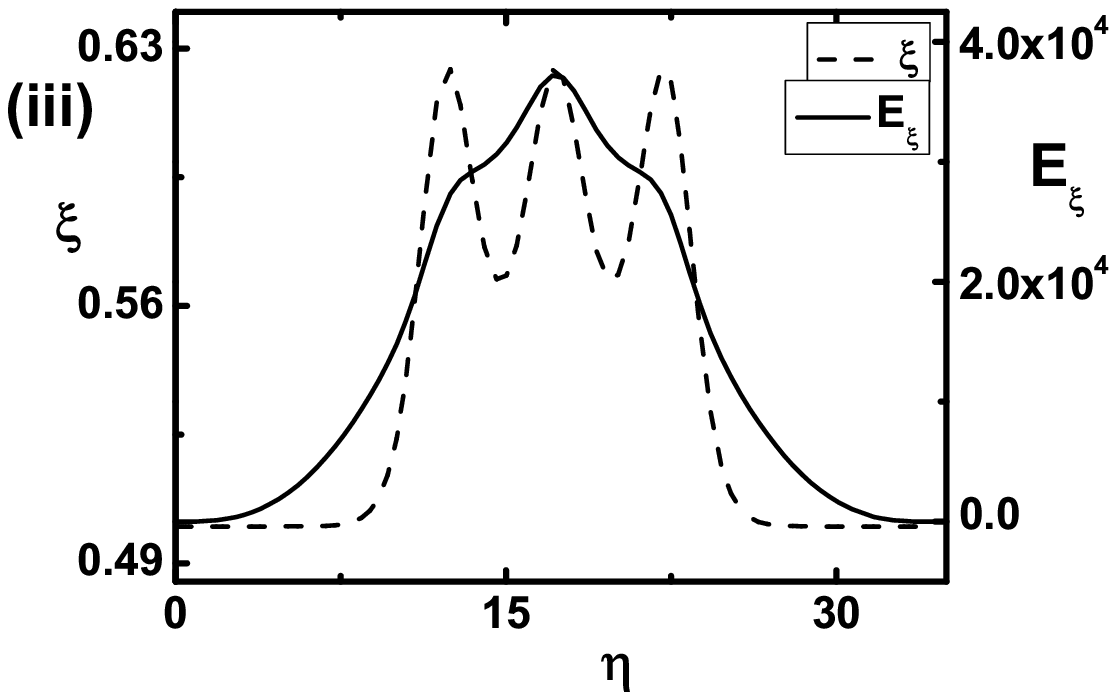}
\includegraphics[width=2.0in]{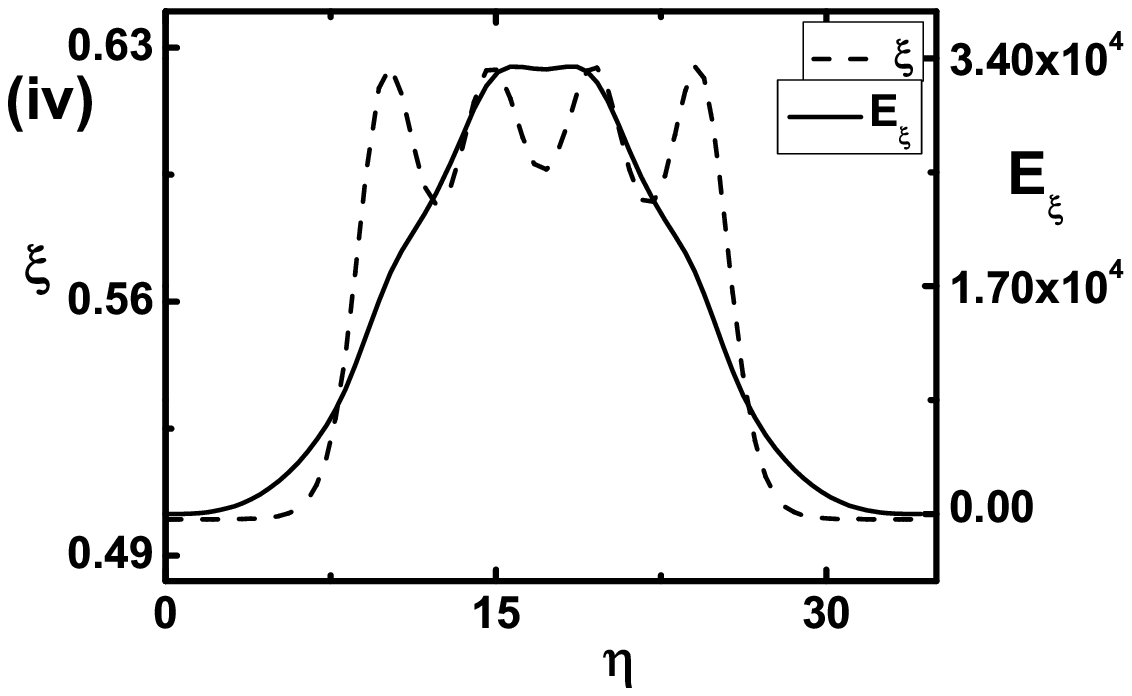}
\includegraphics[width=2.0in]{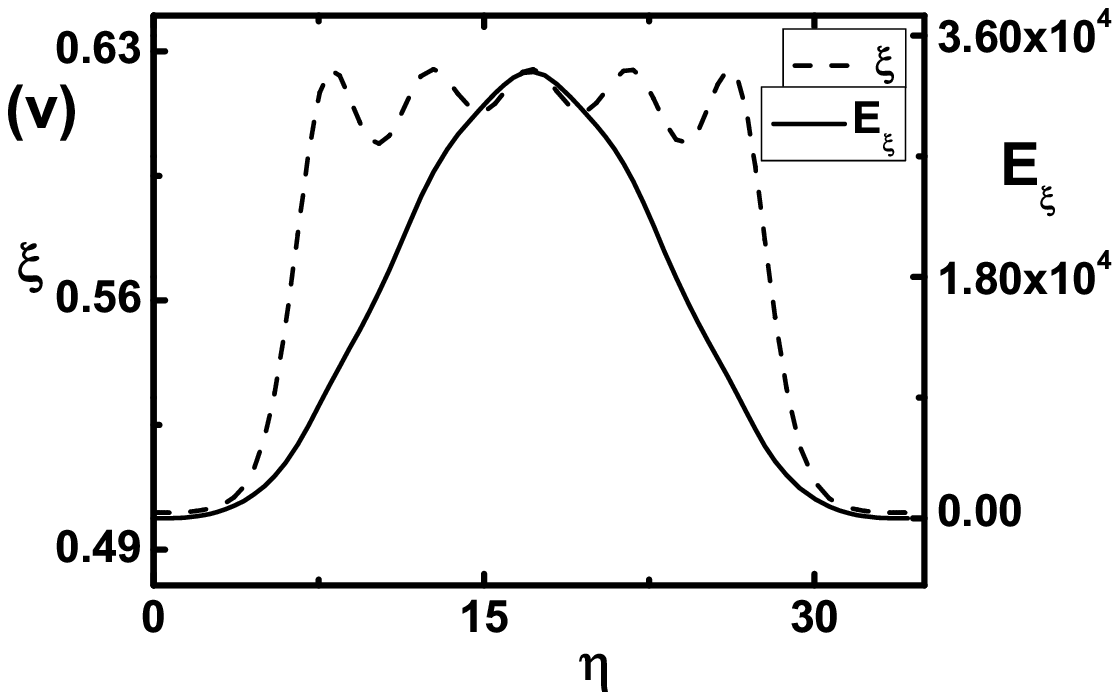}
\includegraphics[width=2.0in]{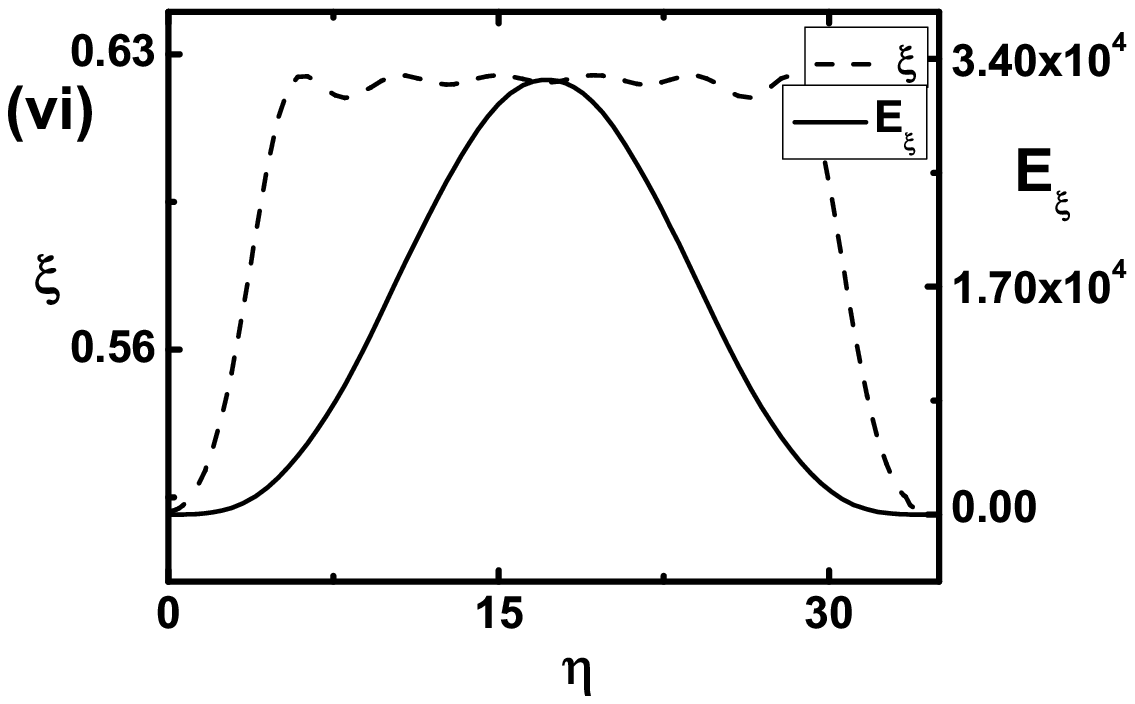}
\end{center}
\caption{\footnotesize{The diagram for energy density of each solutions. In each figure, the dotted line denotes the volume energy density $\xi$ and the solid line denotes the Euclidean energy $E_{\xi}$ at constant $\eta$. }} \label{fig:fig04}
\end{figure}

Figure \ref{fig:fig04} shows the diagram of the energy density for each solutions. In each figure part, the solid line denotes the Euclidean energy $E_{\xi}$ and the dotted line denotes the volume energy density $\xi$ in Eq.\ (\ref{voden}). The Euclidean energy signifies the value after the integration of variables except for $\eta$ in the present case. The peaks represent a rolling phase in the valley of the inverted potential. The maximum value $\xi_{max}$ is equivalent to $U_{top}$. The number of peaks is thus equal to the number of crossing. The peaks broaden in their range near $U_{top}$ as the number of crossing increases. The Euclidean energy also has peaks. However, the shape of the peaks becomes smooth and broadens as the number of crossing increases. As can be seen from figure (vi), the thickness of the wall increases as the number of oscillations increases. The mountain-shaped graph of the Euclidean energy in each figure part is due to the integration of variables in the dS space.

\begin{figure}[t]
\begin{center}
\includegraphics[width=2.0in]{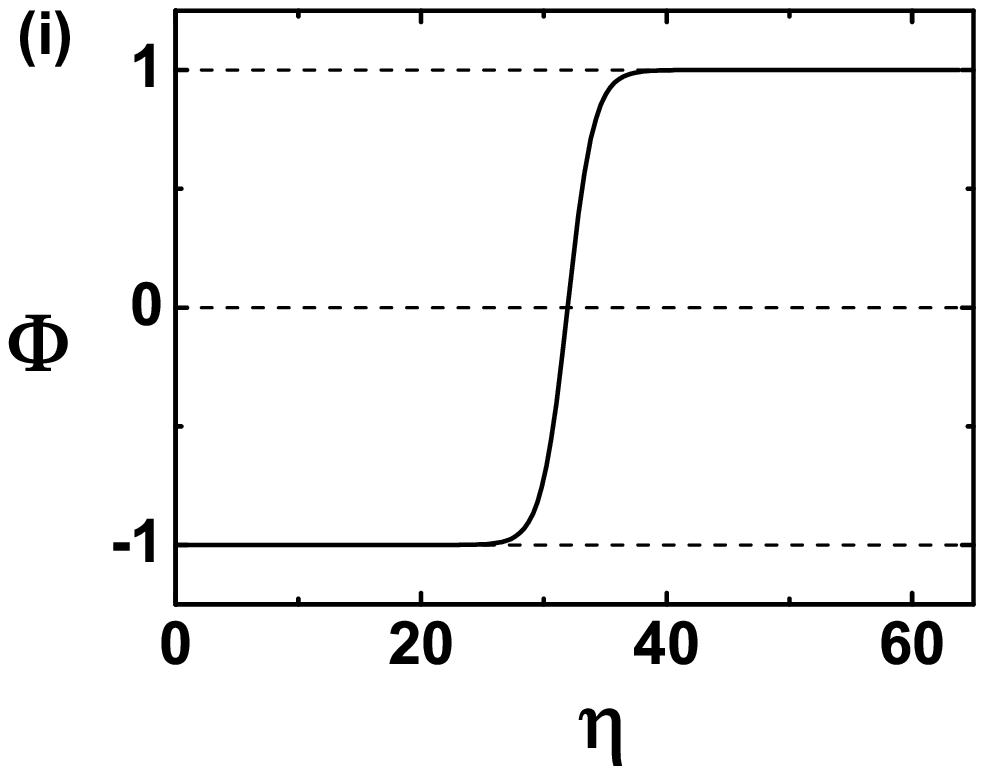}
\includegraphics[width=2.0in]{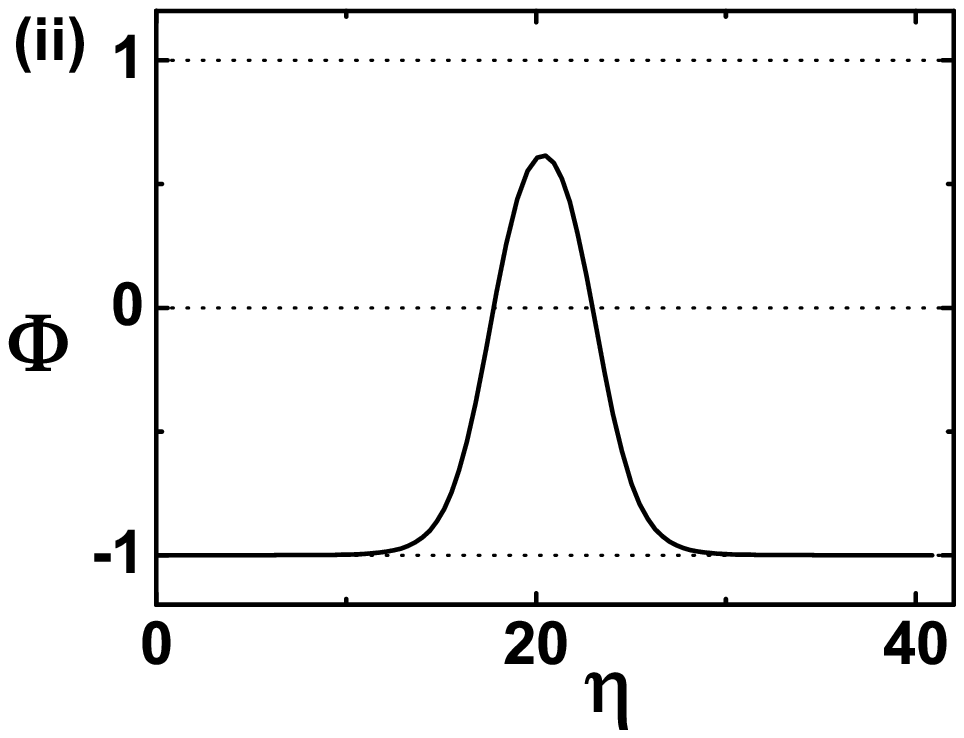}
\includegraphics[width=2.0in]{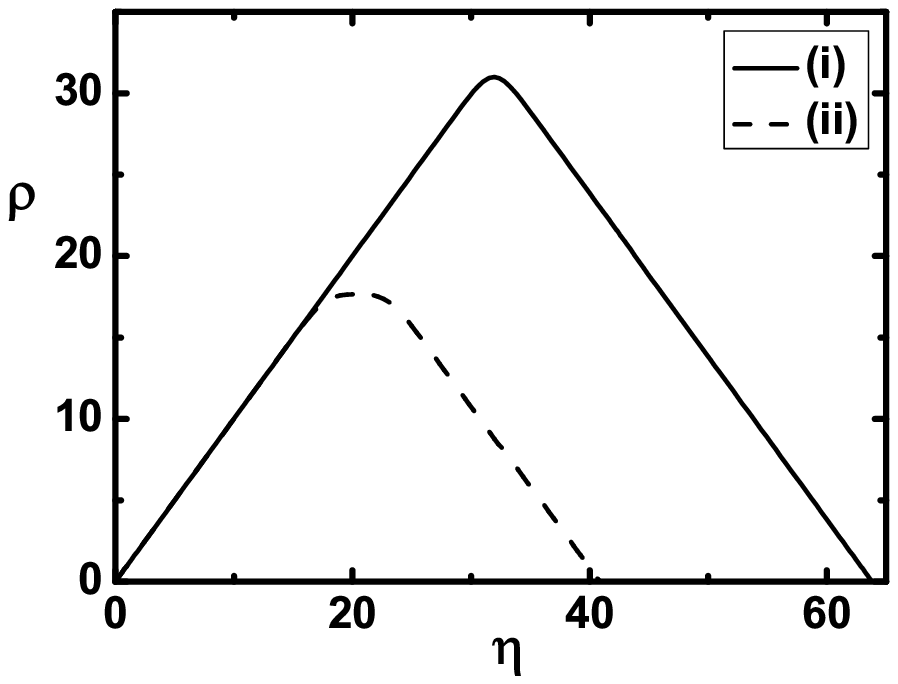}
\end{center}
\caption{\footnotesize{The numerical solutions between flat-flat
degenerate vacua.}}
\label{fig:fig05}
\end{figure}

\begin{figure}[h]
\begin{center}
\includegraphics[width=2.0in]{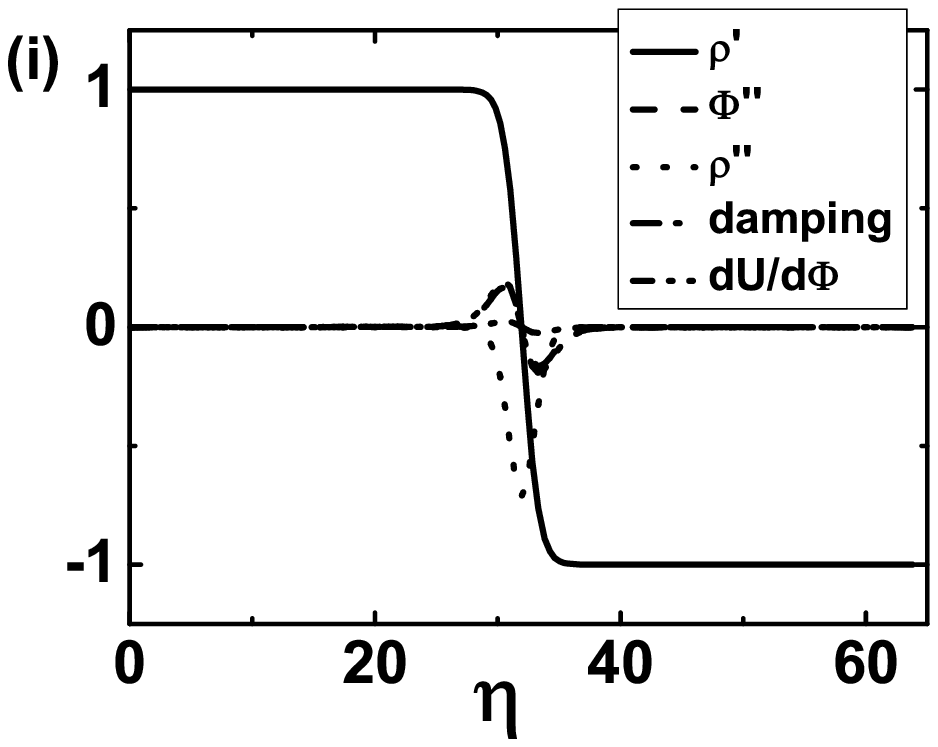}
\includegraphics[width=2.0in]{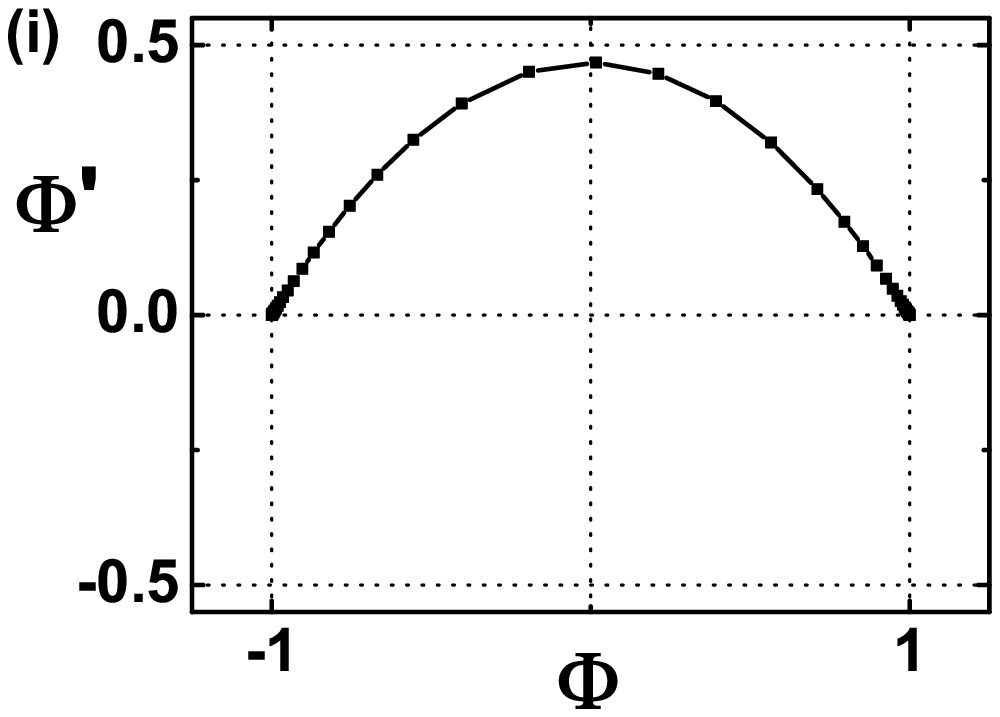}
\includegraphics[width=2.3in]{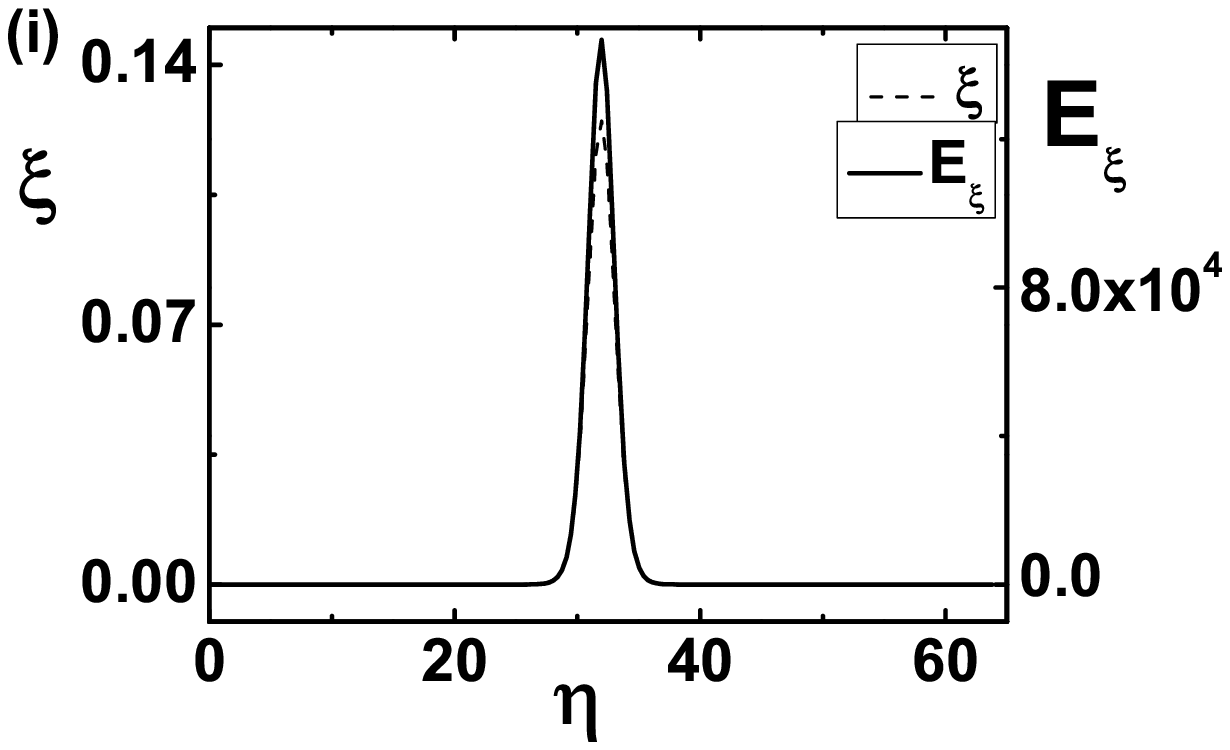} \\
\includegraphics[width=2.0in]{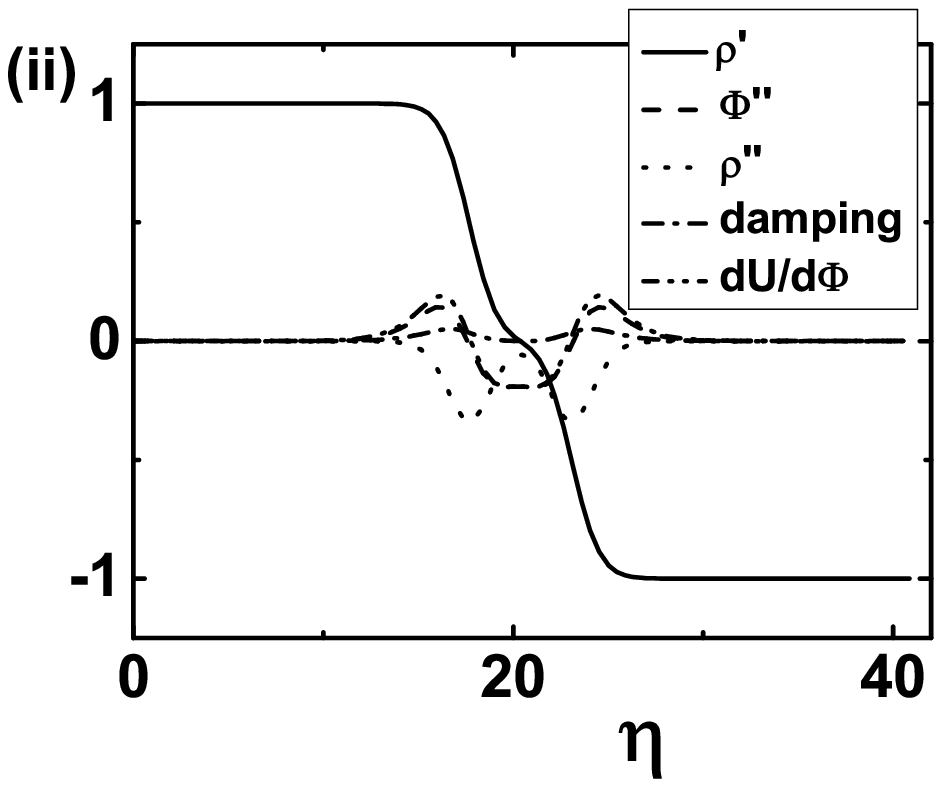}
\includegraphics[width=2.0in]{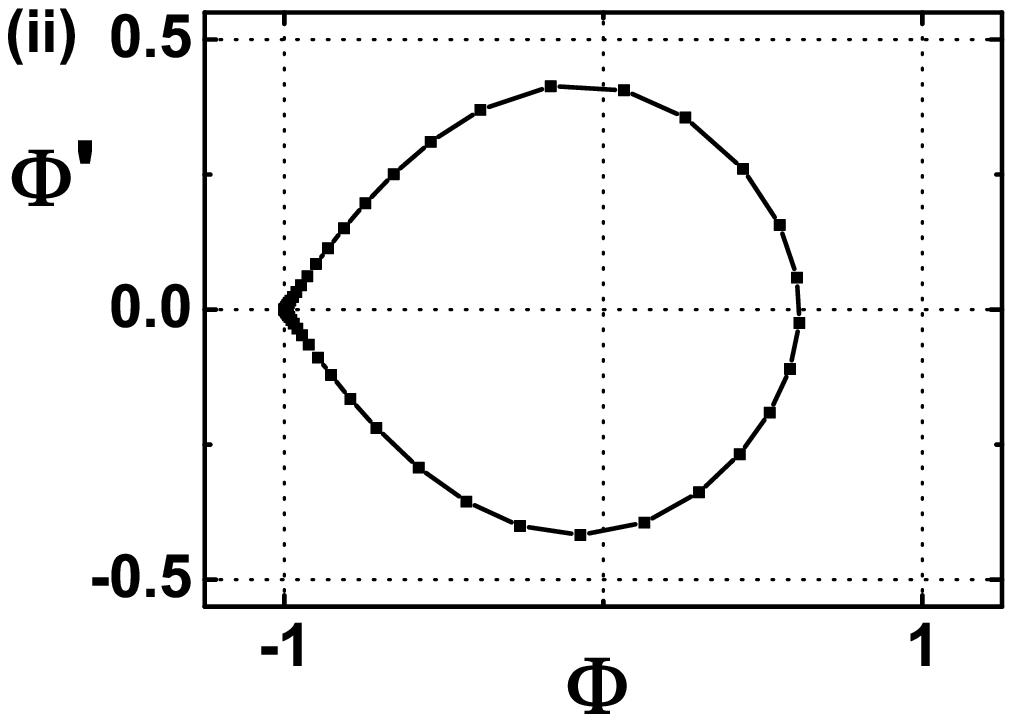}
\includegraphics[width=2.3in]{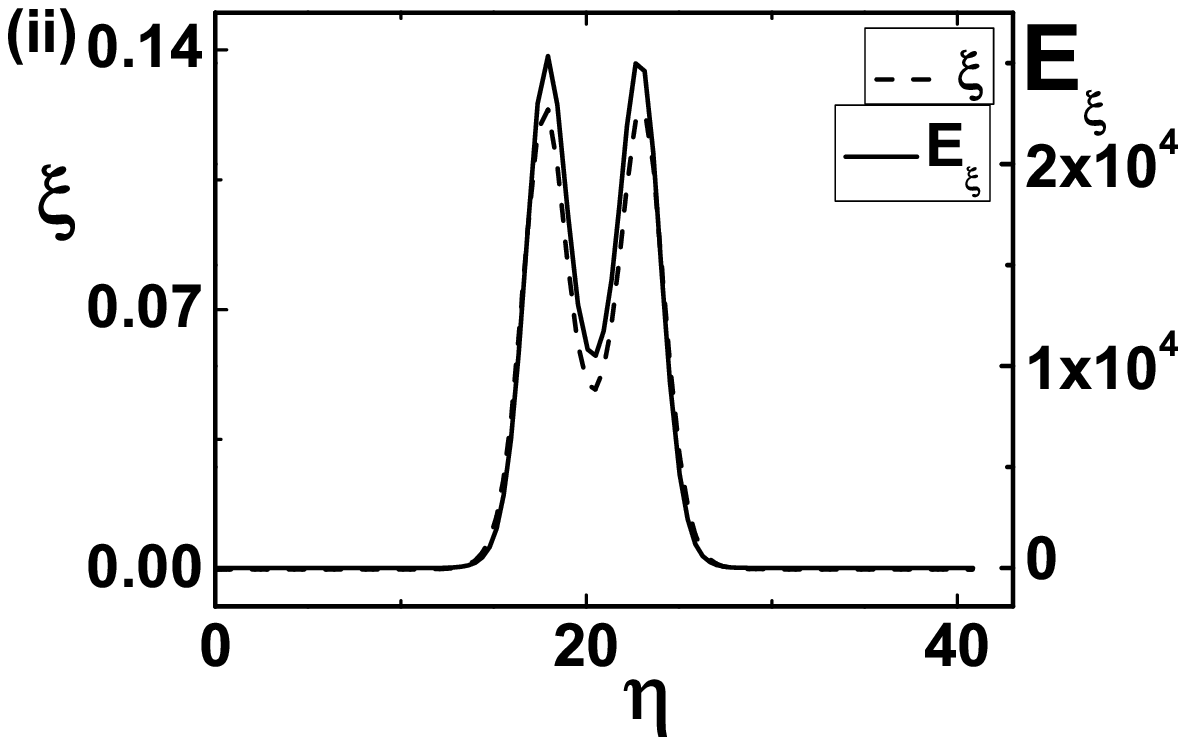}
\end{center}
\caption{\footnotesize{Variation of terms in equations of motion, phase diagrams, and the diagram for energy density between flat-flat degenerate vacua.}}
\label{fig:fig06}
\end{figure}

We now consider oscillating instanton solutions between flat-flat degenerate vacua with $\tilde{U}_o = 0$. The numerical solutions in this case are shown in Fig.\ \ref{fig:fig05}, where we take $\tilde{\kappa}=0.2$ for all the figures. The figure (i) illustrates the solution of $\tilde{\Phi}$ for a one-crossing solution and the figure (ii) for a two-crossing solution. The final figure illustrates the solution for each $\tilde{\rho}$. The solution of $\tilde{\rho}$ is $\eta$ in fixed flat space. Thus, the graph of the linear function near the vacuum states indicates flat space. The maximum number $n_{max}$ is $6$ for the parameters $\tilde{U}_o =0$ and $\tilde{\kappa}=0.2$.

Figure \ref{fig:fig06} depicts a collection of diagrams including the variation of terms in the equations of motion, phase diagrams, and the diagram of energy density between flat-flat degenerate vacuum states. Figure (i) corresponds to a one-crossing solution and figure (ii) corresponds to a two-crossing solution. In the second figure, the maximum value of $\tilde{\Phi}'$ is about $0.49$. The volume energy density and the Euclidean energy are always positive. The initial and final region of $\tilde{\rho}'$ in the first and fourth figures exhibit a horizontal nature as the solution near vacuum states indicates flat space. In Figs.\ \ref{fig:fig05} and \ref{fig:fig06}, two solutions with $n=1$ and $n=2$, are illustrated and other solutions are omitted because the general behaviors for these cases are similar to those in dS space.

\begin{figure}[t]
\begin{center}
\includegraphics[width=2.0in]{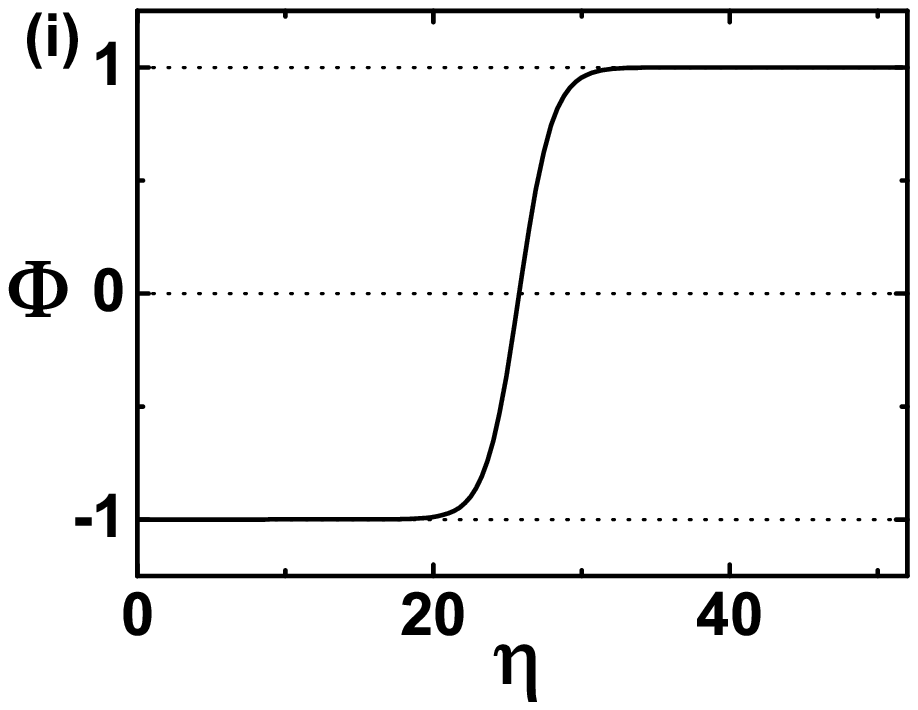}
\includegraphics[width=2.0in]{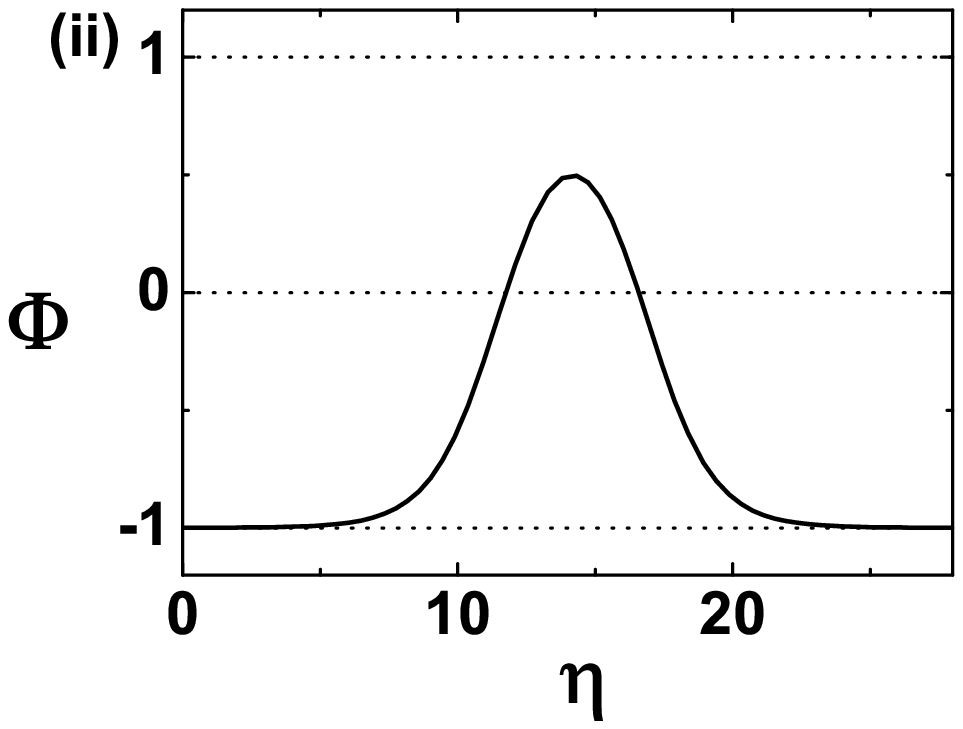}
\includegraphics[width=2.0in]{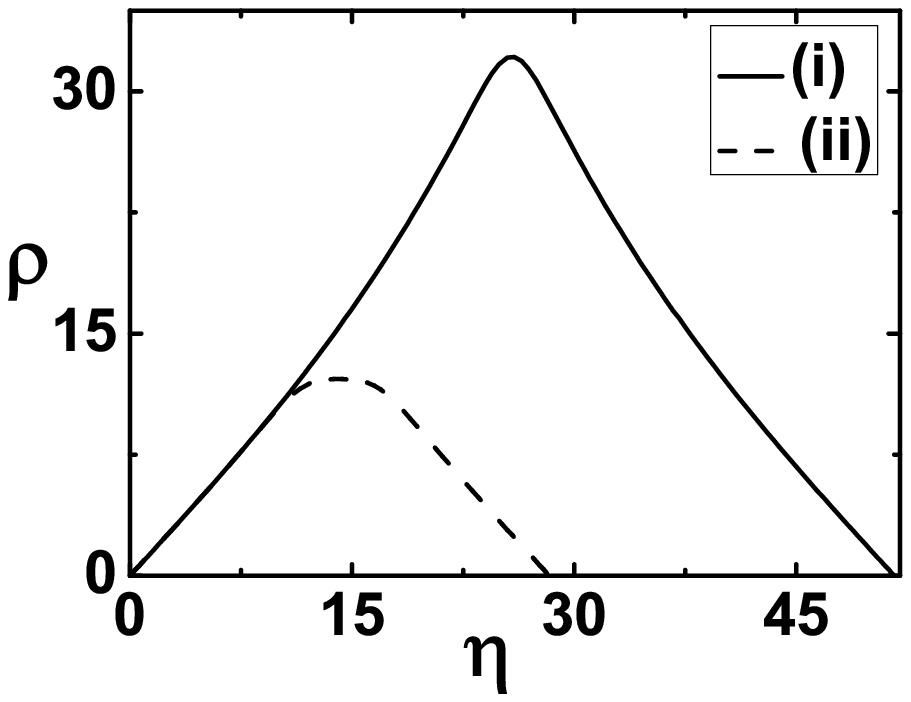}
\end{center}
\caption{\footnotesize{The numerical solutions between AdS-AdS
degenerate vacua.}}
\label{fig:fig07}
\end{figure}

Next, we consider oscillating instanton solutions between AdS-AdS degenerate vacua with $\tilde{U}_o < 0$. The numerical solutions in this case are shown in Fig.\ \ref{fig:fig07}, where we take  $\tilde{U}_o = -0.02$ and $\tilde{\kappa} = 0.4$ for all the figures. Figure (i) illustrates the solution of $\tilde{\Phi}$ for a one-crossing solution while figure (ii) does the same for a two-crossing solution. The third figure illustrates the solution for each $\tilde{\rho}$. The solution of $\tilde{\rho}$ is $\sqrt{\frac{3}{\tilde{\kappa}|\tilde{U}_o|}} \sinh\sqrt{\frac{\tilde{\kappa}|\tilde{U}_o|}{3}}\eta$ in fixed AdS space. Thus, a graph of hyperbolic sine type function near the vacuum states indicates AdS space. The maximum number $n_{max}$ is $4$ for these parameters $\tilde{U}_o$ and $\tilde{\kappa}$.

\begin{figure}
\begin{center}
\includegraphics[width=2.0in]{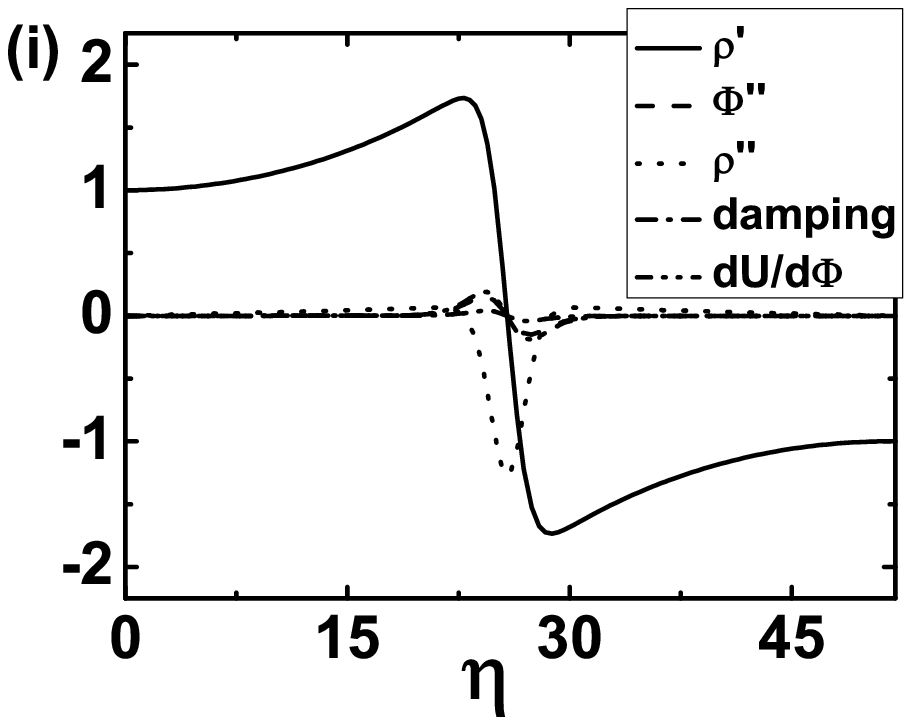}
\includegraphics[width=2.0in]{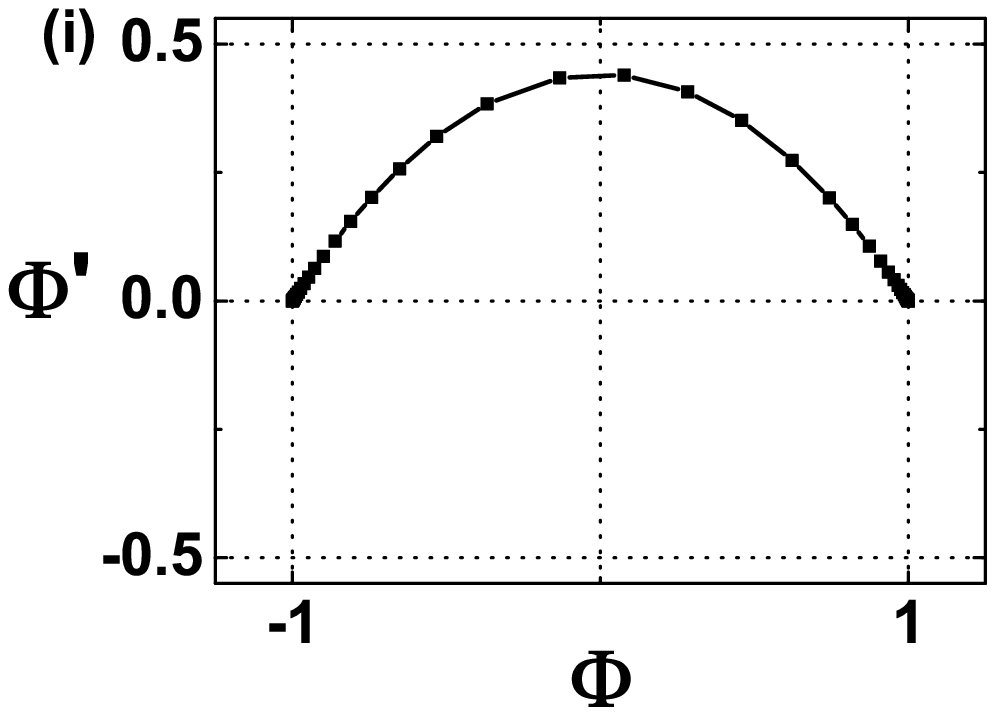}
\includegraphics[width=2.3in]{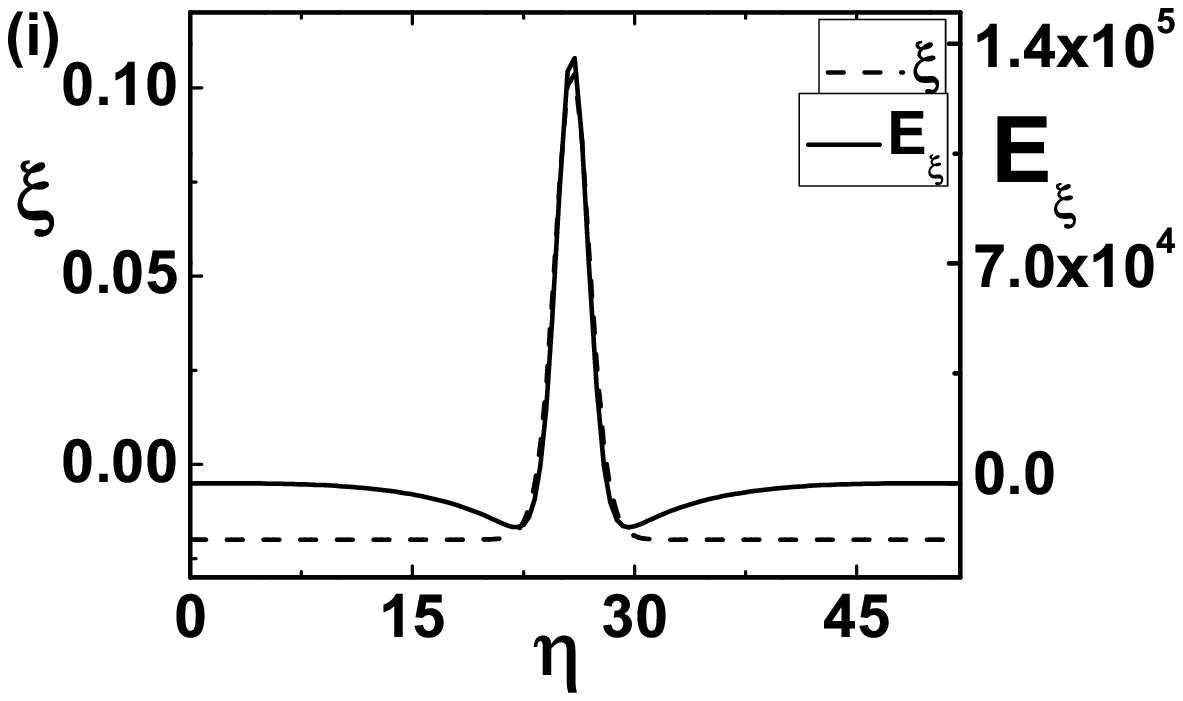}\\
\includegraphics[width=2.0in]{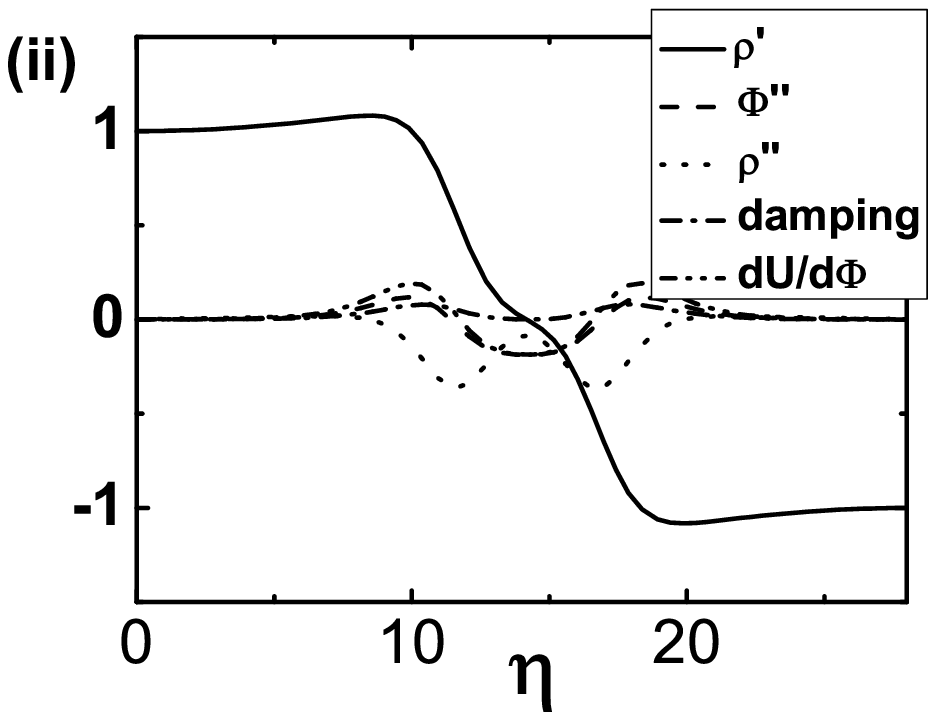}
\includegraphics[width=2.0in]{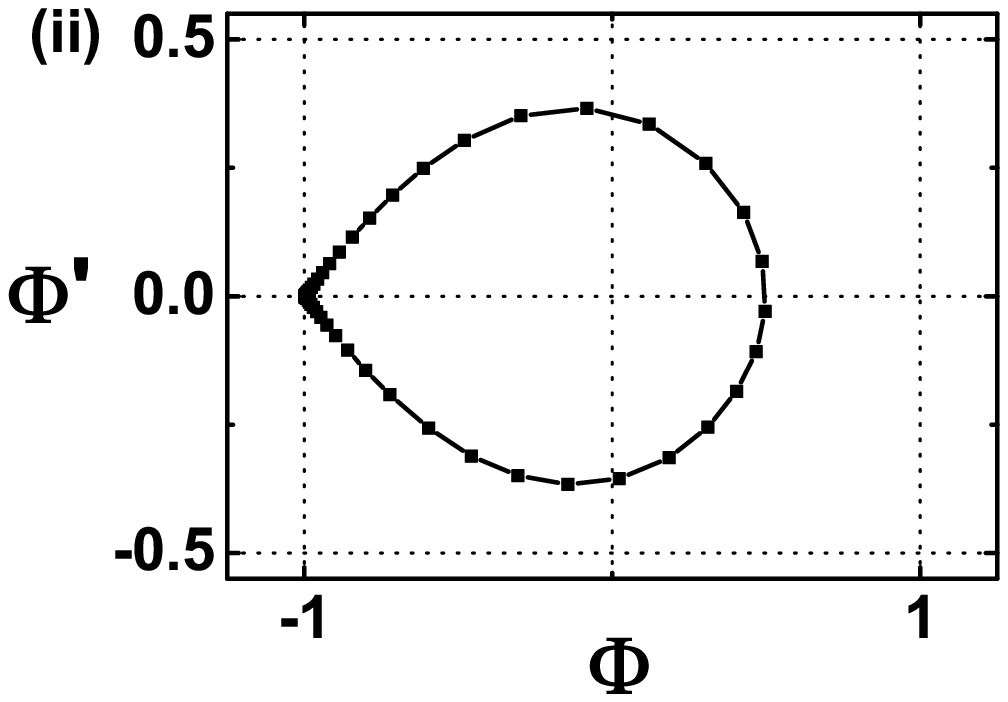}
\includegraphics[width=2.3in]{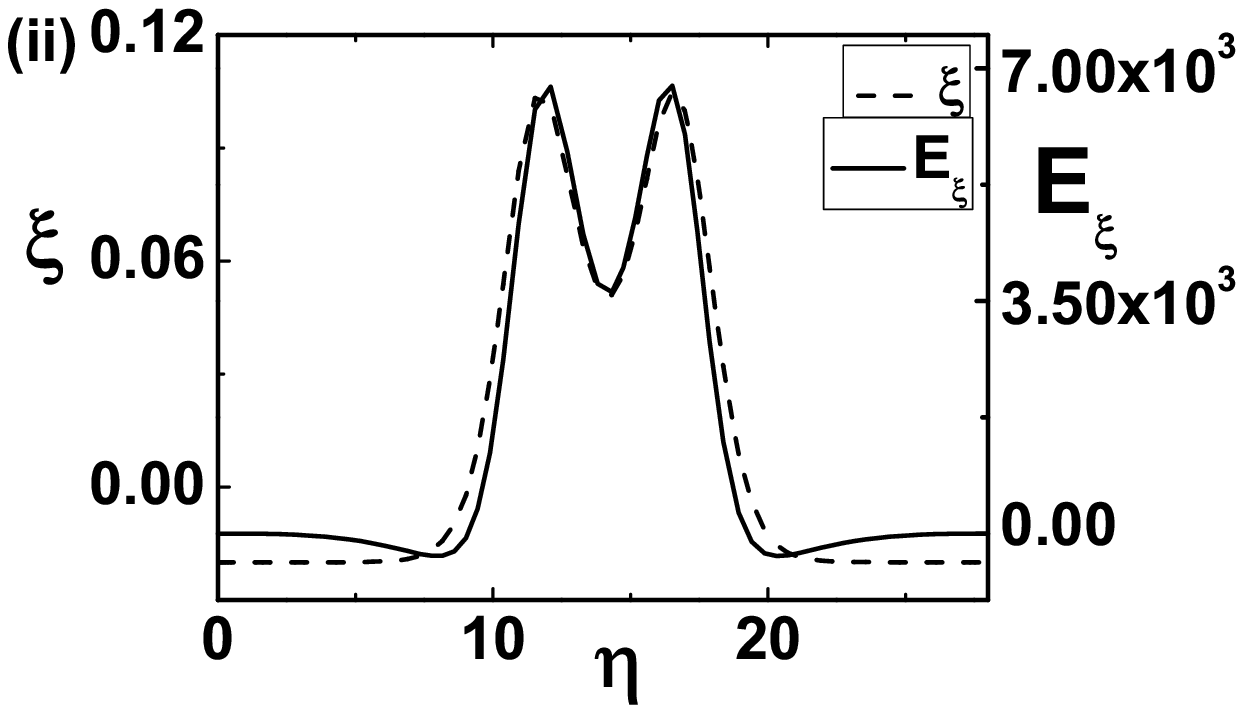}
\end{center}
\caption{\footnotesize{Variation of terms in equations of motion, phase diagrams, and the diagram for energy density between AdS-AdS degenerate vacua.}} \label{fig:fig08}
\end{figure}

Figure \ref{fig:fig08} shows the collective variation of terms in equations of motion, phase diagrams, and the diagram of energy density between AdS-AdS degenerate vacuum states. Figure (i) corresponds to a one-crossing solution while figure (ii) corresponds to a two-crossing solution. In the third and sixth figures, the dented region between the peaks of the Euclidean energy is due to the integration of variables in AdS space. The volume energy density and the Euclidean energy are both negative in AdS region and become positive in the valley of the inverted potential, i.e. the dS region. The initial and the final region of $\tilde{\rho}'$ in the first and fourth figures exhibit a hyperbolic cosine type function, because the solution near the vacuum states indicates AdS space. In Figs.\ \ref{fig:fig07} and \ref{fig:fig08}, two solutions with $n=1$ and $n=2$, are illustrated and all other solutions are omitted since the general behaviors for these cases are similar to those in dS space.

The second term of the first equation in Eq.\ (\ref{erho}) contributes significantly to obtain solutions. In other words, the sign of the term needs to be properly changed to keep harmony with the sign of the other terms. This adequate change of sign is possible only if the gravity is taken into account. As a result of tunneling, geometries of solutions become finite with $Z_2$ symmetry. Each space surrounded by the wall can be either dS, flat, or AdS space. Thus the geometries of solutions in flat and AdS space become quite different from those before tunneling.

We now examine the tunneling probability for the solutions. The tunneling rate can be evaluated in terms of the classical configuration and represented as $e^{-\Delta S}$ in this approximation, where the leading semiclassical exponent $\Delta S=S^{cs} - S^{bg}$ is the difference between the Euclidean action corresponding to a classical solution $S^{cs}$ and the background action $S^{bg}$. When one considers the usual bounce solution, the outside geometry of the bounce solution does not change ever after tunneling. Hence the contribution from outside is canceled. The only contribution to $\Delta S$ comes from the part inside and the wall leading to the finiteness of $\Delta S$.

On the other hand, if one considers a single instanton solution with a finite size, there exists a subtle problem. There is no such subtlety for the dS case. For this case, both the actions for solutions with a finite size of the lens type geometry and background with finite size $S^4$ are negative, i.e. $S^{cs} < 0$ and $S^{bg} < 0$, while the net difference is positive $\Delta S > 0$ giving rise to the probability of less than $1$. However, there is a problem in defining the probability for the flat and AdS space. For flat space the Euclidean action for solution is a negative one, i.e. $S^{cs} < 0$ while the background $S^{bg} =0$. Therefore the net difference is negative $\Delta S < 0$ making the naive probability greater than $1$. The negative value of the exponent is due to the contribution for the valley of the inverted potential, i.e. the dS region, which is related to the fact that the Euclidean action for Einstein gravity is not bounded from below \cite{ghp000}. For AdS space the Euclidean action of solutions is a negative one $S^{cs} < 0$, whereas the background $S^{bg} = + \infty$. The net difference has got the negative divergence $\Delta S = - \infty$. This divergence is due to the contribution coming from the infinite space with nonzero vacuum energy. It is not clear how to interpret this point in physical terms in the present work. We can make $\Delta S$ finite by introducing a cutoff. The effect of the cutoff, as the size increases, was discussed in Ref.\ \cite{lllo}. In this work, we will concentrate only on the relative probability. We then only need to compare the actions for oscillating solutions with the action of a one-crossing instanton, i.e. $(S^{cs}_n - S^{bg}) - (S^{cs}_1 - S^{bg})= S^{cs}_n - S^{cs}_1$.

Table \ref{table1} shows the actions for all of the solutions. We numerically obtained the actions for oscillating solutions from the data. In the integration, we employ an additional normalization process as we use dimensionless quantities for our numerical calculation. Thus we have normalized $\Delta \tilde{S}_{n}$ to be the difference between the action of the $n$-crossing solution and that of a one-crossing solution divided by a $\left| \tilde{S}_{1} \right|$ as follows:
\begin{equation}
\Delta\tilde{S}_{n} \equiv \frac{\Delta \tilde{S}_{n}}{\left| \tilde{S}_{1} \right|}
= \frac{\tilde{S}_{n}-\tilde{S}_{1}}{\left| \tilde{S}_{1} \right|}
\end{equation}
where $\tilde{S}_{n}$ denotes the action for the $n$-crossing solution. Our results are depicted in Table \ref{table1}, in which $\left(\tilde{\kappa}, \tilde{U}_o \right)$ are taken to be (0.04,0.5), (0.2,0), and (0.4,-0.02) respectively. Our results
show that, as predicted, the transition amplitude is suppressed with an increase in oscillation number.

\begin{table}[t]
\caption{\footnotesize{$\Delta S_{n}$ for Oscillating solutions
between dS-dS, flat-flat, and AdS-AdS degenerate vacua.}}
\begin{center}
\renewcommand{\arraystretch}{1.5}
\begin{tabular}{cccc}
  \noalign{\hrule height0.8pt}
  Oscillation number & dS-dS case($\Delta\tilde{S}_{n}$)($e^{-\Delta\tilde{S}_{n}}$) & flat-flat($\Delta\tilde{S}_{n}$)($e^{-\Delta\tilde{S}_{n}}$) & AdS-AdS($\Delta\tilde{S}_{n}$)($e^{-\Delta\tilde{S}_{n}}$) \\
  \hline
  1 & -544000(0)(1) & -403752(0)(1) & -152000(0)(1) \\
  2 & -511000(0.061)(0.941) & -161000(0.601)(0.548) & -39300(0.903)(0.405) \\
  3 & -490000(0.099)(0.906) & -114726(0.716)(0.489) & -30234(0.925)(0.396) \\
  4 & -479000(0.119)(0.887) & -101000(0.750)(0.472) & -28384(0.930)(0.395) \\
  5 & -475000(0.127)(0.881) & -96000(0.762)(0.467) & - \\
  6 & -473000(0.131)(0.878) & -94700(0.765)(0.465) & - \\
  \noalign{\hrule height0.8pt}
\end{tabular} \label{table1}
\end{center}
\end{table}

\section{Properties of oscillating instanton solutions between the degenerate vacua}  \label{sec4}

In the previous section, we obtained various properties of the oscillating solutions between dS-dS, flat-flat, and AdS-AdS degenerate vacua. We make an attempt to obtain the phase space of these solutions in terms of the parameters $\tilde\kappa$ and $\tilde{\kappa}\tilde{U}_o$. The parameter $\tilde\kappa$ is the ratio between the gravitational constant or Planck mass and the mass scale in the theory, $\tilde{\kappa}=\frac{\mu^2}{\lambda}\kappa =\frac{8\pi\mu^2}{M^{2}_{p}\lambda}$, whereas the parameter $\tilde{\kappa}\tilde{U}_o$ is related to the cosmological constant $\Lambda/\mu^2$. Many questions naturally arise at this stage. The first one is how many of the oscillating solutions are allowed for given $\tilde\kappa$ and $\tilde{U}_o$. The second one is whether the number of oscillations depends on these parameters or not. What is the whole phase space of solutions according to the parameter regions, if the number depends on the parameters? The third one deals with the different behaviors of solutions among dS-dS, flat-flat, and AdS-AdS cases. To answer these questions, we will try to figure out the phase space of solutions in terms of $\tilde\kappa$ and $\tilde{\kappa}\tilde{U}_o$. For $\tilde{\kappa} \ll 1$, the effect of gravity is negligible. However, when $\tilde{\kappa}$ approaches to order one value, the effect of gravity becomes important. To obtain the phase space of solutions, we collect data of $\tilde{U}_o$, which makes a difference in the maximum number of oscillations, $n_{max}$, for a given $\tilde{\kappa}$. We then obtain a specific $\tilde{U}_o$ which determines the minimum number of oscillations, $n_{min}$. Finally, we employ the method of least squares \cite{NR} to obtain the relationship between the two parameters in the given data sets of $\tilde\kappa$ and $\tilde{\kappa}\tilde{U}_o$. Oscillating numbers appear to be linearly related to the parameters.

\begin{figure}[t]
\begin{center}
\includegraphics[width=2.5in]{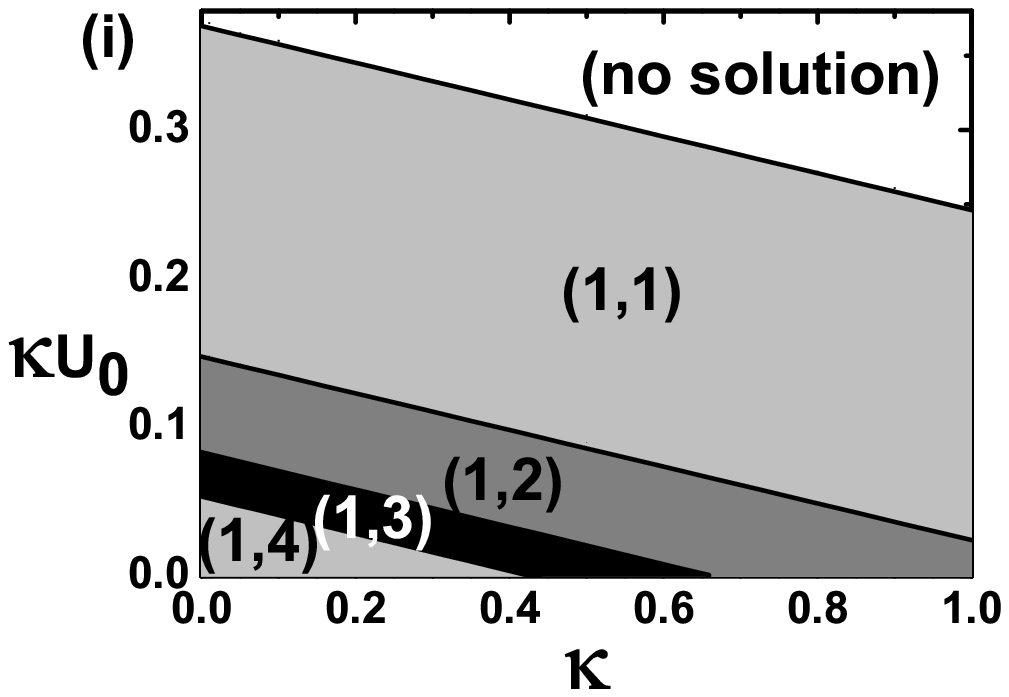}
\includegraphics[width=2.5in]{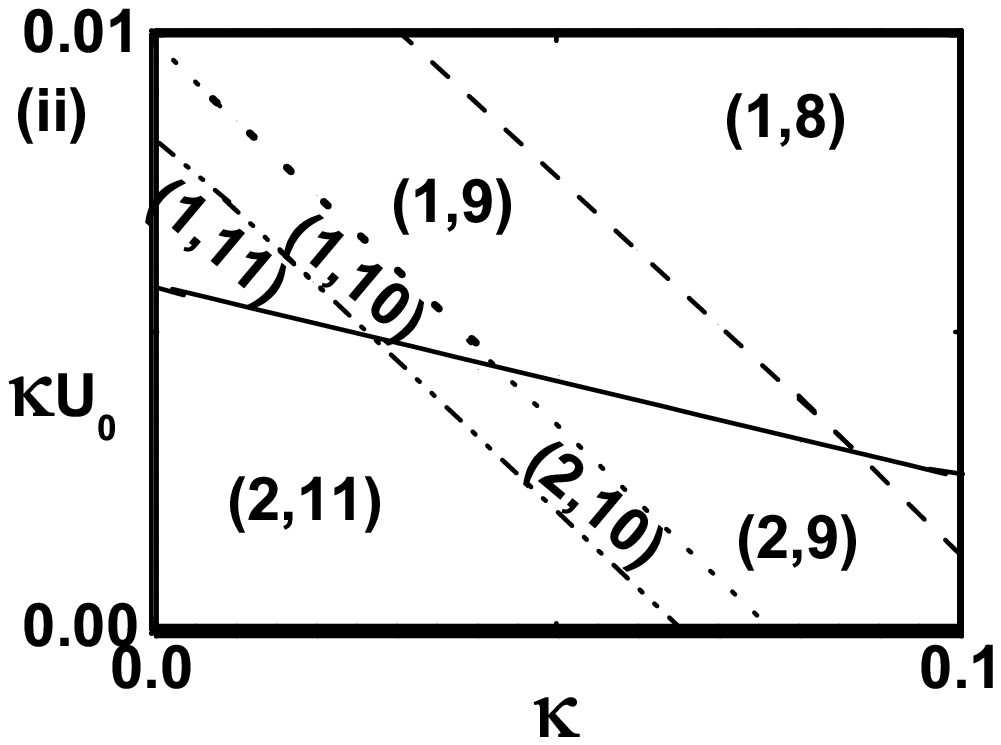} \\
\includegraphics[width=2.5in]{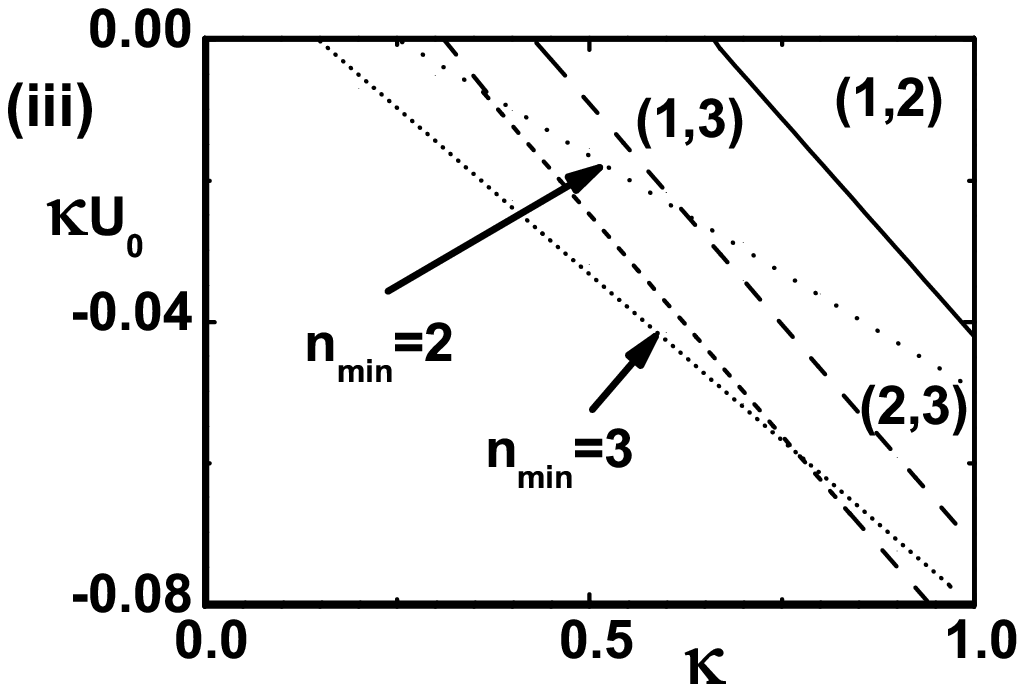}
\includegraphics[width=2.5in]{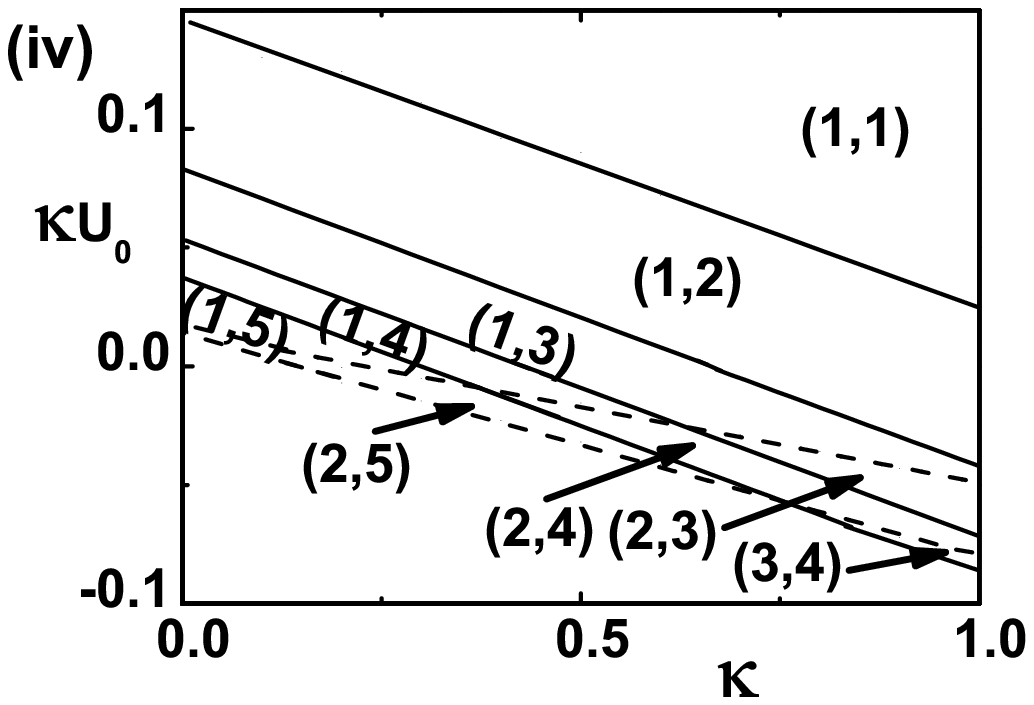}
\end{center}
\caption{\footnotesize{Oscillating behavior depending on $\tilde{\kappa}$ and
$\tilde{\kappa}\tilde{U}_o$. Figure(i) shows the oscillating properties of dS-dS
degenerate case and figure(ii) shows small values of parameters once
more. Figure(iii) is a AdS-AdS degenerate case, and figure(iv) shows the total parameter space, we have searched.}} \label{fig:fig09}
\end{figure}

Figure \ref{fig:fig09} shows the behavior of oscillating solutions in terms of $\tilde{\kappa}$ and $\tilde{\kappa}\tilde{U}_o$. The value of $\tilde{\kappa}$ is limited in the range $0 \leqq \tilde{\kappa} \leqq 1$. At the point $\tilde{\kappa}=0$, or $M_p = \infty$, gravity is switched off. The $y$ axis represents no gravity. We expect that there exists no solution with $O(4)$ symmetry when gravity is turned off. In each of the figure parts, we use the notation ($n_{min}$, $n_{max}$), where $n_{min}$ represents the minimum number of oscillations and $n_{max}$ the maximum number of oscillations in the given parameter range. For example, ($n_{min}$, $n_{max}$) = ($1$, $1$) means that the minimum number of oscillations is $1$ and the maximum number of oscillations is $1$. Figures (i) and (ii) illustrate the number of oscillations for the case of dS-dS degenerate vacua. In fig.\ (i), there is a zone representing no solution in the upper right region. When the parameters $\tilde{\kappa}$ or $\tilde{U}_o$ are increased, the evolution period of $\tilde{\eta}_{max}$ is diminished in general. $\tilde{\rho}$ approaches $\tilde{\rho}_{max}$ before the field arrives at the other vacuum state in the inverted potential. Thus, there is no solution because of the short evolution period. In other words, the instanton solution can not fit inside the Euclideanized dS background in the strong gravity limit. When the parameters $\tilde{\kappa}$ or $\tilde{U}_o$ are decreased, both the evolution period of $\tilde{\eta}_{max}$ and the maximum number of oscillations get increased. Figure (ii) illustrates the zone with small value of parameters in fig.\ (i). The upper zone allows $n_{min}=1$, whereas the lower zone allows $n_{min}=2$ divided by the solid line. Figure (iii) illustrates the behavior of the oscillating solutions in the case of AdS-AdS degenerate vacua. In this case, $(n_{min},n_{max})=(1,1)$ is not allowed. If there is a tunneling solution, then there are oscillating solutions found together. This is a different property as compared to the dS-dS case. In addition to this, there is a no-solution parameter region corresponding to regions for which $\tilde{U}_o \leq -0.125$ in the case of AdS-AdS degenerate vacua as there is no dS region of potential \cite{lllo}. In this case, the change of $n_{min}$ is more prominent. Even at $\tilde{\kappa} = 1$, with strong gravity, the change of $n_{min}$ occurs. This is another difference between the dS-dS and the AdS-AdS cases. Whole parameter space of $\tilde{\kappa}$ and $\tilde{\kappa}\tilde{U}_o$ is shown in the fig.\ (iv). In this figure, we can see that the changes of $n_{min}$ and $n_{max}$ in the dS-dS case are continuously connected to those in the AdS-AdS case. The $x$ axis, $\tilde{U}_o=0$, represents the flat-flat case. In the flat-flat case, there exists a solution for all of $\tilde{\kappa}$ except for $\tilde{\kappa}=0$.

\begin{figure}[t]
\begin{center}
\includegraphics[width=3.0in]{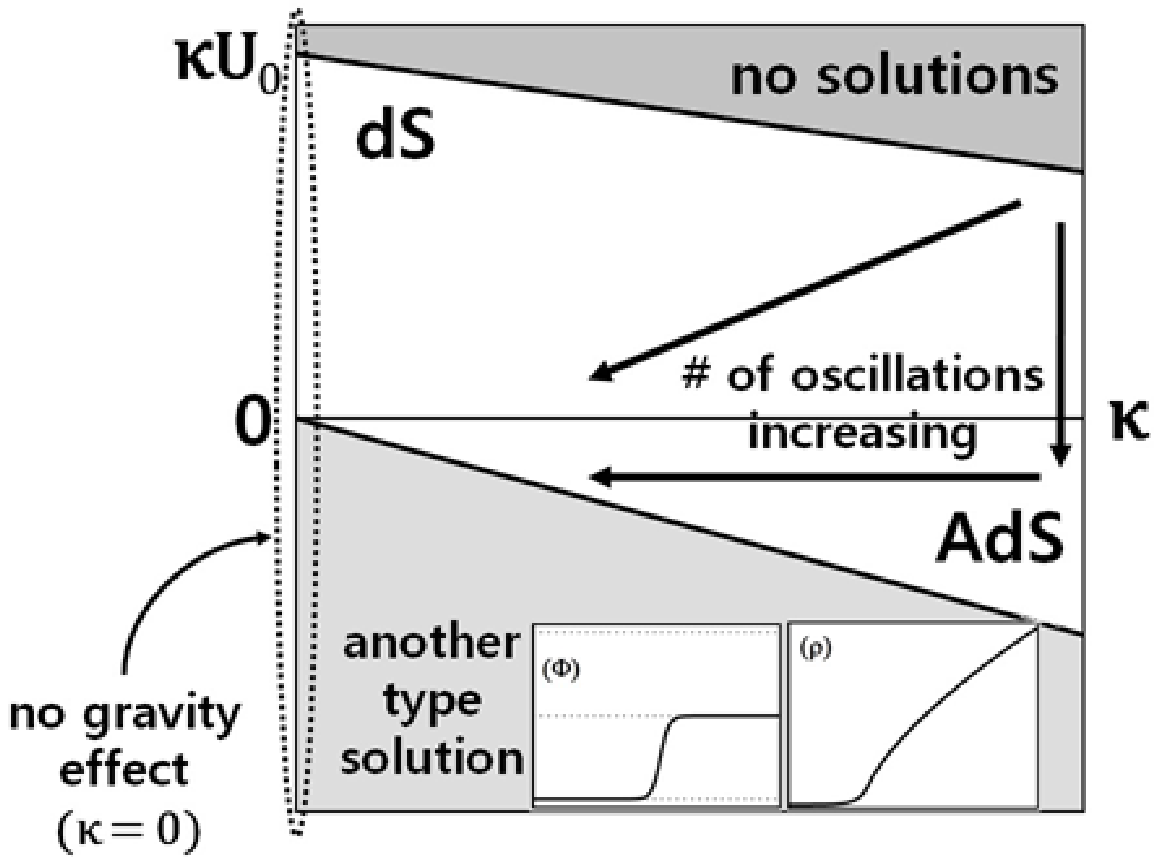}
\includegraphics[width=3.0in]{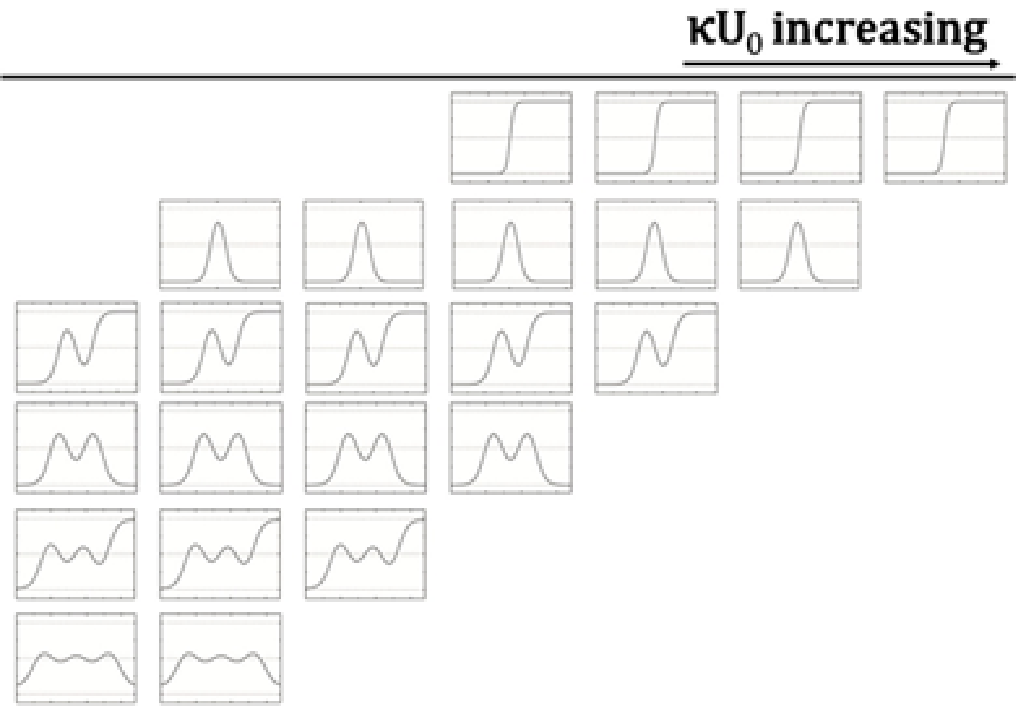}
\end{center}
\caption{\footnotesize{Illustrations of the phase space of solutions and oscillating properties. Left figure shows a whole solution space, and right figure shows number of oscillating solutions changing on
given $\kappa$.}} \label{fig:fig10}
\end{figure}

Figure \ref{fig:fig10} shows the schematic diagram for the phase space of all solutions including yet another type of solution and the number of oscillating solutions with different $\tilde{\kappa}$ values. The left figure has $\tilde{\kappa} = 0$ line indicating no gravity effect. There is a zone representing no solution in the upper right region in the case of dS. The dS region has positive $\tilde{U}_o$ and the AdS region has negative $\tilde{U}_o$ divided by the line, $\tilde{U}_o=0$, representing the flat case. In the middle area including the flat case, $n_{min}$ and $n_{max}$ increase as $\tilde{\kappa}$ and $\tilde{\kappa}\tilde{U}_o$ decrease. The tendency is indicated as the painted arrows. The inclined line in the AdS region represents $\tilde{U}_o=-0.125$ or $\tilde{U}_{top}=0$. In the lower left region, there exists another type of solution. The figure on the right shows how $n_{min}$ and $n_{max}$ are changed in terms of $\tilde{\kappa}\tilde{U}_o$ and $\tilde{\kappa}$. As we can see from the figure, $n_{max}$ and $n_{min}$ increase as $\tilde{\kappa}\tilde{U}_o$ decreases.

\begin{figure}[t]
\begin{center}
\includegraphics[width=2.0in]{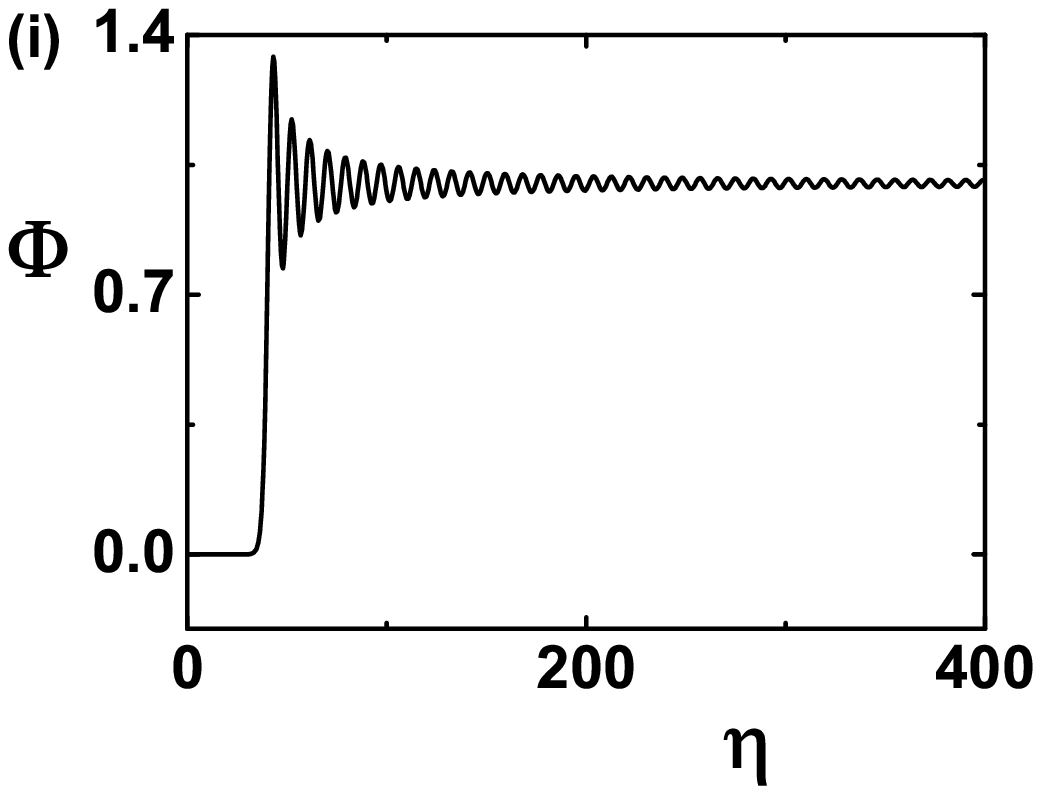}
\includegraphics[width=2.0in]{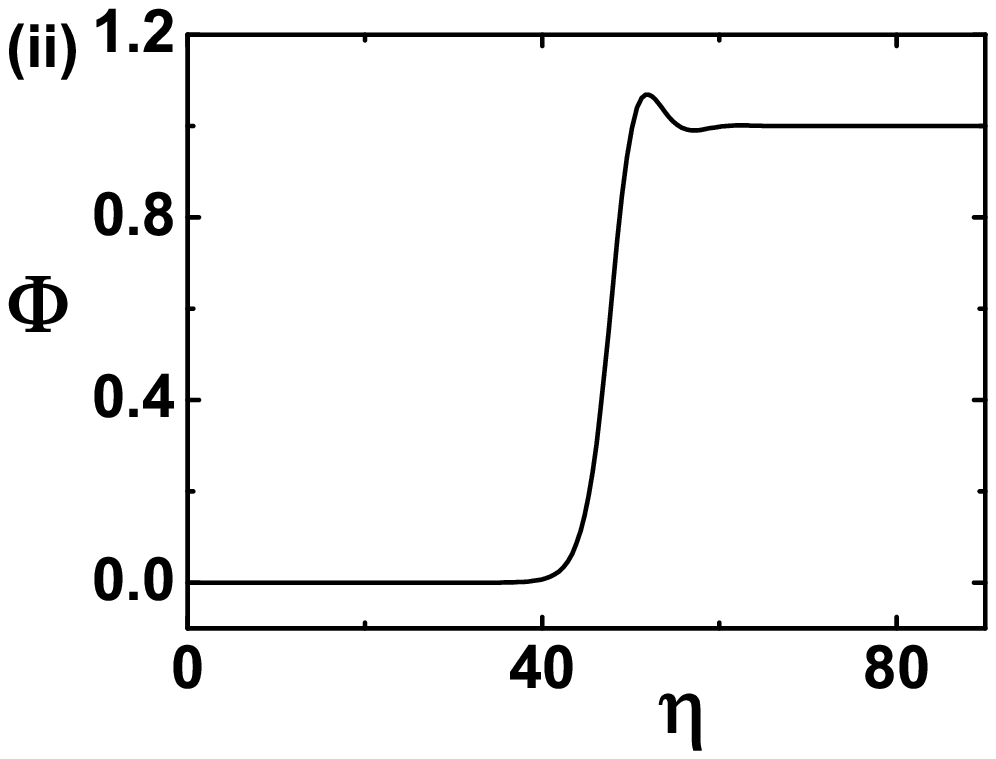}\\
\includegraphics[width=2.0in]{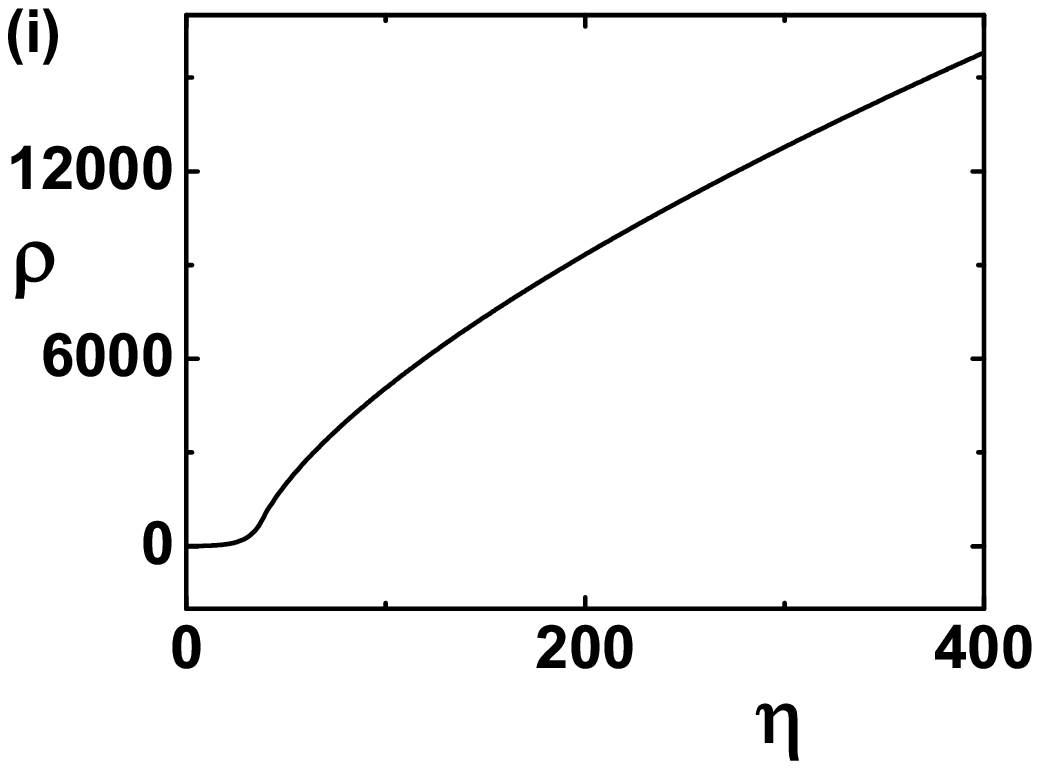}
\includegraphics[width=2.0in]{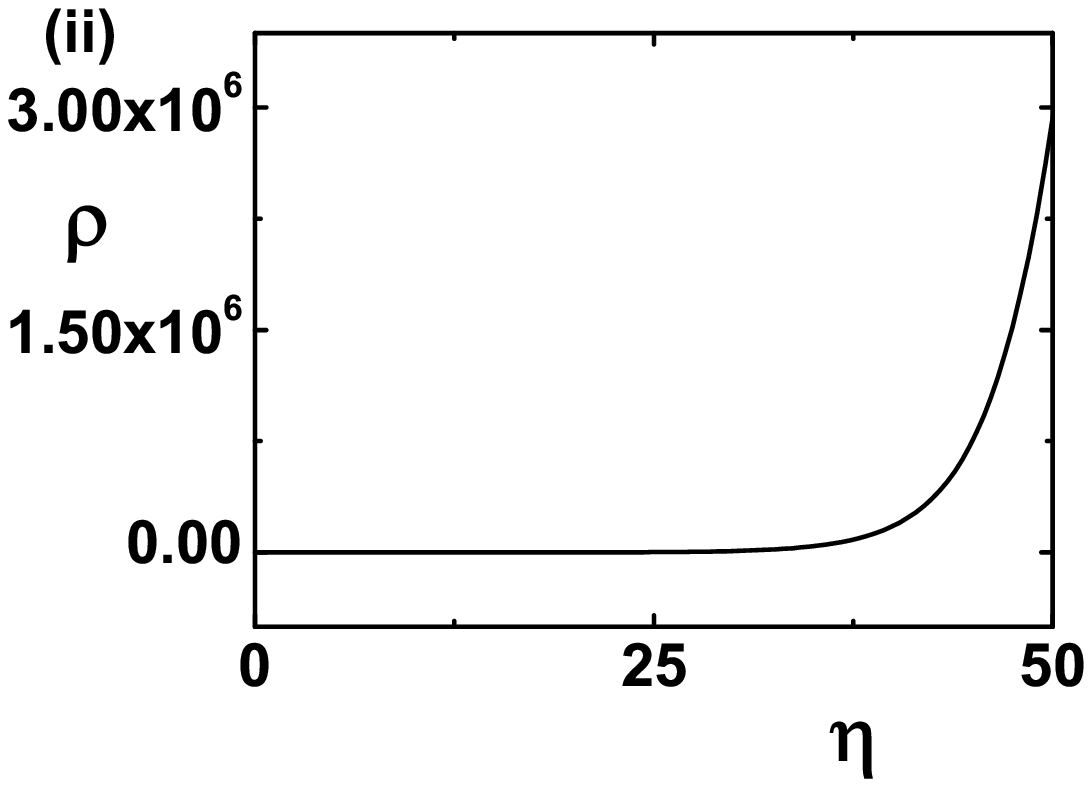}
\end{center}
\caption{\footnotesize{The solutions describing tunneling without a barrier. The upper two figures show the solution of $\tilde{\Phi}$ and the lower two figures show the solution of $\tilde{\rho}$. }} \label{fig:fig11}
\end{figure}

Figure \ref{fig:fig11} shows another type of solutions in Fig.\ \ref{fig:fig10}. The solutions represent tunneling from the top of the potential, a point of an unstable equilibrium, to the local vacuum state instead of rolling down the potential. The solutions can be of the same kind of a bubble solution describing tunneling without a barrier \cite{klee08, ljps}. In figures (i), we take $\tilde{\kappa}=0.5$ and $\tilde{U}_o=-0.125$. The inside geometry is AdS and that of the outside is flat. In figures (ii), we take $\tilde{\kappa}=0.5$ and $\tilde{U}_o=-1$. The inside geometry is AdS and that of the outside is AdS.

\section{Summary and Discussions}

We have studied oscillating instanton solutions of a self-gravitating scalar field between degenerate vacua. We obtained numerical solutions in dS background. Basically, our method for obtaining the domain wall or braneworld-like object \cite{rs10, pdm00} is based on the instanton-induced theory rather than the kink-induced theory. Our approach is related to the question: How we can make spacetime including the domain wall in dS and flat space? If the thin-wall approximation scheme is allowed in our work and the mechanism is applied to solutions in  higher-dimensional theories, our one-crossing solutions can be interpreted as the mechanism providing nucleation of the domain wall or braneworld in instanton-induced theory. Because our oscillating solutions have a thick wall with varying energy density, our oscillating solutions can be interpreted as the mechanism providing nucleation of the thick wall for topological inflation \cite{avil04, linde07}. We add that $Z_{2}$ invariant solutions also exist in flat or AdS background, though the physical significance is not clear. Furthermore, we constructed the phase space of all our solutions including the number of oscillations. As a by-product, we obtained the solution describing tunneling without a barrier.

In Sec.\ II, we analyzed the boundary conditions of our problem. There are two kinds of conditions. We imposed the boundary conditions as the initial value problem and imposed additional conditions implicitly. The initial value $\tilde{\Phi}_o$ is obtained by employing the undershoot-overshoot procedure. To avoid a singular solution at $\tilde{\eta} \rightarrow \tilde{\eta}_{max}$ in Eq.\ (\ref{eqrho-change}) and demand the $Z_2$ symmetry, the conditions $d\tilde{\Phi}/{d\tilde{\eta}} \rightarrow 0$ and $\tilde{\rho}\rightarrow 0$ as $\tilde{\eta}\rightarrow \tilde{\eta}_{max}$ were needed.

In Sec.\ III, we obtained the numerical solution of oscillating instantons. In particular, we performed the numerical work with more detail for the cases in dS background. The number of oscillations increases as the initial point, $\tilde{\Phi}(\tilde{\eta}_{initial}) =\tilde{\Phi}_o$, moves away from the vacuum state. We can see that the size of the geometry of solutions decreases as the number of crossing increases because the period of the evolution parameter $\tilde{\eta}_{max}$ decreases as the starting point moves away from the vacuum state. As expected, this type of solutions is possible only if gravity is taken into account. The maximum number of oscillations is determined by the parameters $\tilde{\kappa}$ and $\tilde{U}_o$ observed in Ref.\ \cite{hw000}. The solutions are of two types. One type is for situations in which different parts of spacetime are in different vacua. The solutions representing tunneling which go from the left vacuum state to the right vacuum state have a nonzero topological charge. The other type is for situations in which different parts of spacetime are in the same vacuum state. Others representing the solutions which go back to the starting point after oscillations have zero topological charge. We can see that the sign change of $\tilde{\rho}'$ from positive to negative occurs at the half period of $\tilde{\eta}$ due to the $Z_2$ symmetry. We have studied the behavior of the solutions in the $\tilde{\Phi}(\tilde{\eta})$ -$\tilde{\Phi}'(\tilde{\eta})$ plane using the phase diagram method. Figures \ref{fig:fig03}(i), (iii), and (v) have symmetry about the $y$ axis, while figures \ref{fig:fig03}(ii), (iv), and (vi) have symmetry about the $x$ axis. The maximum value of $\tilde{\Phi}'$ decreases as the number of crossing increases. We have checked the energy density diagrams for each solutions. The peaks of the volume energy density broaden in range near $U_{top}$ as the number of crossing increases. The shape of peaks of the Euclidean energy becomes smooth and broadens as the number of crossing increases.

We have examined the tunneling probabilities for the solutions. The tunneling rate can be evaluated in terms of the classical configuration and can be represented as $e^{-\Delta S}$ in this approximation, where the leading semiclassical exponent $\Delta S$ is the difference between the Euclidean action corresponding to a classical solution and the background action. When one considers the usual bounce solution, the outside geometry of the bounce solution does not change even after tunneling. Hence the contribution from the outside gets canceled. The only contribution to $\Delta S$ stems from the inside part and the wall leading to the finiteness of $\Delta S$.

On the other hand, if one considers a single instanton solution with finite size, there exist a subtle problem. There is no such subtlety for the dS case. However, there is a problem in defining the probability for the flat and the AdS background. The Euclidean action for the flat space case is a negative one, $\Delta S < 0$ making the naive probability greater than $1$. The Euclidean action for the case of AdS space $\Delta S \rightarrow - \infty$. This negative divergence of the exponent is due to the contribution coming from the infinite space with nonzero vacuum energy. It is yet not clear how to interpret this point correctly in the present work. Of course, we can make $\Delta S$ finite by introducing a cutoff. The effect of the cutoff, as the size increases, was discussed in Ref.\ \cite{lllo}. In the present work, we concentrate mainly on the relative probability. We only need to compare the action of oscillating solutions with that of a one-crossing instanton. Our results show that, as predicted, the transition amplitude is suppressed with an increase in the number of oscillations. One may concern the contribution from the boundary term \cite{York}. We expect that the point $\tilde{\eta} = \tilde{\eta}_{max}$ in the opposite side after the tunneling process is smooth due to $Z_2$ symmetry and therefore the boundary term does not contribute to the action \cite{lllo}.

In Sec.\ IV, we obtained the phase space of solutions in terms of the parameters $\tilde\kappa$ and $\tilde{\kappa}\tilde{U}_o$. To make the phase space of solutions, we collected data on $\tilde{U}_o$, which makes the difference of the maximum number of oscillations, $n_{max}$, for a given $\tilde{\kappa}$. We then obtained a specific $\tilde{U}_o$ which determines the minimum number of oscillations, $n_{min}$. Finally, we employed the method of least squares \cite{NR} to obtain the relationship between two parameters in given data sets of $\tilde\kappa$ and $\tilde{\kappa}\tilde{U}_o$. Oscillating numbers appeared to be linearly related to the parameters.
Figure \ref{fig:fig09} shows the behaviors of oscillating solutions in terms of $\tilde{\kappa}$ and $\tilde{\kappa}\tilde{U}_o$. The value of $\tilde{\kappa}$ is limited to $0 \leqq \tilde{\kappa} \leqq 1$. At the point $\tilde{\kappa}=0$, or $M_p = \infty$, gravity is switched off. The $y$ axis represents no gravity. We expect that there is no solution with $O(4)$ symmetry when gravity is switched off. When the parameter $\tilde{\kappa}$ or $\tilde{U}_o$ increased, the evolution period of $\tilde{\eta}_{max}$ diminished in general.  When the parameter $\tilde{\kappa}$ or $\tilde{U}_o$ decreased, the evolution period of $\tilde{\eta}_{max}$ and the maximum number of oscillations increased. There is a no-solution parameter region corresponding to regions in which $\tilde{U}_o \leq -0.125$ in the case of AdS-AdS degenerate vacua because there is no dS region of potential \cite{lllo}. We can see the changes of $n_{min}$ and $n_{max}$ in the dS-dS are continuously connected to those in the AdS-AdS. The $x$ axis, $\tilde{U}_o=0$, represents the flat-flat case. In the flat-flat case, there exists a solution for all of $\tilde{\kappa}$ except when $\tilde{\kappa}=0$. Figure \ref{fig:fig10} shows the schematic diagram of the phase space of all solutions including another type solution and the number of oscillating solutions with different $\tilde{\kappa}$s.

As a result of this tunneling, a finite-sized geometry with $Z_2$ symmetry is obtained. Our mechanism for making the domain wall or braneworld-like object is different from the ordinary formation mechanism of the domain wall because our solutions are instanton solutions rather than soliton solutions. In other words, our solutions can be interpreted as solutions describing an instanton-induced domain wall rather than a kink-induced domain wall or braneworld-like object. Domain walls can form in any model having a spontaneously broken discrete symmetry. An inertial observer sees the domain wall accelerating away with a specific acceleration. The domain wall has repulsive gravitational fields \cite{dvil, dip}. The thickness of the domain wall in flat spacetime can be estimated by a balance between the potential energy and the gradient energy. When the thickness of the domain wall is greater than or equal to the horizon size corresponding to the vacuum energy in the interior of the domain wall, topological inflation can occur. The scalar field stays near the top of the potential at the core. This potential energy serves as a vacuum energy in a similar way to the slow-rollover inflationary models. This topological inflation does not require fine-tuning of the initial conditions and is eternal even at the classical level due to the topological reason. Our oscillating instanton solutions can be interpreted as mechanism providing the nucleation of the thick wall for the topological inflation. The wrinkles representing the variation of the volume energy density in the wall may be interpreted as density perturbations in the inflating region. In this work, inflating regions described by the oscillating solutions and density perturbations described by the variation of energy density can occur simultaneously. Furthermore, oscillating bounce solutions also have the thick wall. Thus we expect that (non-)topological inflation can be made by oscillating bounce solutions.

To obtain the dynamics of the solutions, we should apply the analytic continuation from Euclidean to Lorentzian signature. If the wall is thin, we can employ the Israel junction condition \cite{lllo, isr01, bkt} or the method \cite{bnu02, chw02}. If the wall is thick, the double-null simulation may be more relevant to the dynamics of the solutions \cite{jddy}.

A direction for future research could be whether or not our solutions obtained using the instanton-induced theory can be extended to theories with gauge fields or various other dimensions and whether or not the topological charge of both instanton and bounce solutions, including oscillating one, can be well-defined and conserved for a self-gravitating scalar field in various dimensions. It would also be interesting to examine if this toy model with the variation of energy density can provide a proper inflationary scenario. In Ref.\ \cite{mgll}, the creation of a charged black hole pair separated by a thin domain wall and the dynamics of the domain wall were studied in the cosmological context. It would be interesting to study the properties and evolution of topological inflation with a magnetic field.

\section{Acknowledgements}
We would like to thank E.~J.~Weinberg, Yun Soo Myung, Soonkeon Nam, Hongsu Kim, Jungjai Lee, Gungwon Kang, Youngone Lee, Hyeong-Chan Kim, Hyun Seok Yang, In Yong Park, and Dong-han Yeom for helpful discussions and comments. We would like to thank Remo Ruffini, Hyung Won Lee, and She-Sheng Xue for their hospitality at the 12th Italian-Korean Symposium on Relativistic Astrophysics in Pescara, Italy, 4-8 Jul 2011. This work was supported by the Korea Science and Engineering Foundation (KOSEF) grant funded by the Korea government(MEST) through the Center for Quantum Spacetime(CQUeST) of Sogang University with grant number R11 - 2005 - 021. W.L. was supported by the National Research Foundation of Korea Grant funded by the Korean Government (Ministry of Education, Science and Technology)[NRF-2010-355-C00017].

\end{document}